\documentclass[useAMS,usenatbib]{mn2e}
\usepackage{tikz}
\usepackage{amssymb}

\title[Fragmentation criteria in self-gravitating discs]{On the fragmentation criteria of self-gravitating protoplanetary discs}
\author[Farzana Meru and Matthew R. Bate]{Farzana Meru\thanks{farzana@astro.ex.ac.uk} and Matthew R. Bate\\
School of Physics, University of Exeter, Stocker Road, Exeter, EX4 4QL}
\begin{document}

\maketitle

\label{firstpage}

\begin{abstract}
We investigate the fragmentation criterion in massive self-gravitating discs.  We present new analysis of the fragmentation conditions which we test by carrying out global three-dimensional numerical simulations.  Whilst previous work has placed emphasis on the cooling timescale in units of the orbital timescale, $\beta$, we find that at a given radius the surface mass density (i.e. disc mass and profile) and star mass also play a crucial role in determining whether a disc fragments or not as well as where in the disc fragments form.  We find that for shallow surface mass density profiles ($p<2$, where $\Sigma \propto R^{-p}$), fragments form in the outer regions of the disc.  However for steep surface mass density profiles ($p \gtrsim 2$), fragments form in the inner regions of a disc.  In addition, we also find that the critical value of the cooling timescale in units of the orbital timescale, $\beta_{\rm crit}$, found in previous simulations is only applicable to certain disc surface mass density profiles and for particular disc radii and is not a general rule for all discs.  We find an empirical fragmentation criteria between the cooling timescale in units of the orbital timescale, $\beta$, the surface mass density, the star mass and the radius.

\end{abstract}

\begin{keywords}
accretion, accretion discs - protoplanetary discs - planets and satellites: formation - gravitation - instabilities - hydrodynamics
\end{keywords}

\section{Introduction}
\label{sec:intro}

The evolution of protoplanetary discs has been explored at great length in the past to understand the processes by which they form in the early Class 0 phase, the accretion from the molecular cloud core onto the disc in the Class I phase, the mass and angular momentum transfer in discs once they have formed as well as in the early Class II stage of an isolated disc, through to the disc dispersal mechanisms.  One such concept that has been considered is the importance of the disc self-gravity, particularly in the earlier period of a disc's lifetime.  It is during the early stages when it is massive enough to be self-gravitating that the concepts of angular momentum transport and fragmentation in these discs are important.  This is a particularly significant aspect when considering whether gas giant planets could form via this method.  Historically, planet formation by gravitational instability has not been thought likely since the planets that form in this way are predicted to do so far out in a disc ($ \gtrsim \rm O(100)~AU$ e.g. \citealp{Rafikov_unrealistic_conditions,Matzner_Levin2005,Rafikov_SI,Clarke2009_analytical,Boley_CA_and_GI,Stamatellos_no_frag_inside_40AU}; \citealp*{Kratter_runts}), whereas until recently, giant planets have been found only at small radii ($\lesssim 10$~AU).

Recent advances in observations, such as the discoveries of planets at large radii ($\rm O(10-100)~AU$) from the parent star \citep{Fomalhaut, HR8799} call for an alternative planet formation mechanism other than the standard core accretion method to be considered.  Furthermore, theoretical advances, such as the increased likelihood that planets can form by gravitational instability closer to the parent star in low metallicity environments \citep{Meru_Bate_opacity} where core accretion finds it difficult \citep{Kornet_etal_CA_metallicity}, call for planet formation by gravitational instability to be scrutinised in much more detail.

There are two quantities that have historically been used to determine whether a disc is likely to fragment.  The first is the stability parameter \citep{Toomre_stability1964},

\begin{equation}
  \label{eq:Toomre}
  Q=\frac{c_{\rm s}\kappa_{\rm ep}}{\pi\Sigma G},
\end{equation}
where $c_{\rm s}$ is the sound speed in the disc, $\kappa_{\rm ep}$ is the epicyclic frequency, which for Keplerian discs is approximately equal to the angular frequency, $\Omega$, $\Sigma$ is the surface mass density and $G$ is the gravitational constant.  Therefore, once the surface mass density and the rotation of the disc have been established, the stability is purely dependent on the disc temperature.  \cite{Toomre_stability1964} showed that for an infinitesimally thin disc to fragment, the stability parameter must be less than a critical value, $Q_{\rm crit} \approx 1$.
 
\cite{Gammie_betacool} showed that in addition to the stability criterion above, the disc must cool at a fast enough rate.  Using shearing sheet simulations, he showed that if the cooling timescale can be parametrized as

\begin{equation}
  \label{eq:beta}
  \beta = t_{\rm cool}\Omega,
\end{equation}
where

\begin{equation}
  \label{eq:tcool}
  t_{\rm cool} = u \big(\frac{{\rm d}u_{\rm cool}}{{\rm d}t}\big)^{-1},
\end{equation}
$u$ is the specific internal energy and ${\rm d}u_{\rm cool}/{\rm d}t$ is the total cooling rate, then for fragmentation we require $\beta \lesssim 3$, for a ratio of specific heats $\gamma = 2$ (in two dimensions).  \cite*{Rice_beta_condition} carried out three-dimensional simulations using a Smoothed Particle Hydrodynamics ({\sc sph}) code and showed that this cooling parameter is dependent on the equation of state.  They showed that fragmentation can occur if $\beta < \beta_{\rm crit}$ where $\beta_{\rm crit} \approx 6-7$ for discs with $\gamma = 5/3$ and $\beta_{\rm crit} \sim 12-13$ for discs with $\gamma = 7/5$.  \cite{Gammie_betacool} and \cite{Rice_beta_condition} also showed that in a steady state disc where the dominant form of heating is that due to gravitational instabilities, since the gravitational stress in a disc can be linked to the cooling timescale in the disc using

\begin{equation}
  \label{eq:beta_alpha}
  \alpha_{\rm GI} = \frac{4}{9} \frac{1}{\gamma (\gamma - 1)} \frac{1}{\beta},
\end{equation} 
the rapid cooling required for fragmentation can also be interpreted as a maximum gravitational stress that a disc can support without fragmenting, which they show to be $\alpha_{\rm GI, max} \approx 0.06$.

The concept of a fast cooling needed for fragmentation is very clear from previous work.  However, the value of the critical cooling timescale, $\beta_{\rm crit}$ (and therefore, by equation~\ref{eq:beta_alpha}, $\alpha_{\rm GI,max}$), does not appear to be too clear cut: \cite{Rice_Gammie_confirm} found that for a $0.1 \rm M_{\odot}$ disc with surface mass density profile, $\Sigma \propto R^{-7/4}$, extending to a radius, $R_{\rm out} = 25$~AU around a $1\rm M_{\odot}$ star, the disc fragments using $\beta=3$ but not for $\beta=5$, whereas for a disc with mass $M_{\rm disc} = 0.25 \rm M_{\odot}$, the disc fragments for $\beta=5$.  On the other hand, \cite{Rice_beta_condition} suggest that the fragmentation boundary is independent of the disc to star mass ratio.  \cite*{Libby_MSci} showed that the critical value of $\beta$ (below which fragmentation will occur if the stability criterion is met) may depend on the disc's thermal history: if the timescale on which the disc's cooling timescale is decreased is slower than the cooling timescale itself (i.e. a gradual decrease in $\beta$) then the critical value may decrease by up to a factor of 2.  More recently, \cite*{Cossins_opacity_beta} showed that the critical value varies with the temperature dependence of the cooling law.  In addition, they carry out a simulation of a self-gravitating disc with surface mass density profile, $\Sigma \propto R^{-3/2}$ (c.f. \cite{Rice_beta_condition} who used $\Sigma \propto R^{-1}$), with ratio of specific heats, $\gamma = 5/3$, and show that the critical value $\beta_{\rm crit} \approx 4$.  Using equation~\ref{eq:beta_alpha}, this is equivalent to $\alpha_{\rm GI, max} = 0.1$ which brings the result of  $\alpha_{\rm GI, max} = 0.06$ described above into question.  Yet a number of papers have been produced which base their work on the concept of a single critical value of $\beta$ (or equivalently, a maximum gravitational stress value) \citep[e.g.][]{Clarke2009_analytical,Rafikov_SI,Cossins_opacity_beta,Kratter_runts}.

However, there is a more fundamental question that arises from previous numerical studies using a parametrised cooling method.  In many simulations \citep[e.g.][]{Rice_Gammie_confirm,Rice_beta_condition,Libby_MSci}, the fragments all form in the outer parts of the discs.  If the fragmentation criterion for a self-gravitating disc only depends on $\beta$, then if fragments form, one would expect them to form at all radii since all radii have the same value of $\beta$.  This implies that the cooling timescale is not the only parameter on which fragmentation depends.

In this paper, we investigate the criteria for fragmentation in greater detail.  In Section~\ref{sec:analytics} we analytically investigate how fragmentation may be expected to depend on various disc parameters.  We test this analytical theory by carrying out global three-dimensional simulations, the numerical setup of which we describe in Section~\ref{sec:numerics}.  In Section~\ref{sec:benchmark} we make initial comparisons between our simulations and previous simulations by \cite{Rice_beta_condition} as well as discussing the implications of the disc setup.  In Section~\ref{sec:main_sim}, we present our simulations, the results of which we describe in Section~\ref{sec:results}.  We finally discuss and make conclusions in Sections~\ref{sec:disc} and~\ref{sec:conc}, respectively.

\section{Analytical view}
\label{sec:analytics}
As discussed in Section~\ref{sec:intro}, for a disc to fragment one criteria is that the Toomre stability parameter, $Q \lesssim 1$.  Making the approximation that $\kappa_{\rm ep} \approx \Omega = \sqrt{GM_{\star}/R^3}$ and using  $H = c_{\rm s}/\Omega$, where $H$ is the isothermal scale height of the disc, the Toomre stability criterion becomes a condition on the aspect ratio, $H/R$:

\begin{equation}
  \frac{H}{R} \lesssim \frac{\pi \Sigma R^2}{M_\star}.
  \label{eq:H_R_condition}
\end{equation}
Approximating the surface mass density as $\Sigma \approx M_{\rm disc}/(\pi R^2)$, equation~\ref{eq:H_R_condition} becomes \citep{Gammie_betacool}

\begin{equation}
  \frac{H}{R} \lesssim \frac{M_{\rm disc}}{M_{\star}}.
  \label{eq:analytics_i}
\end{equation}
The surface mass density of a disc can also be written in the form of a power-law, $\Sigma = \Sigma_o(R_o/R)^{p}$, where $\Sigma_o$ is the surface mass density at radius $R_o$, and $p$ is a constant for any one disc.  Substituting this into equation~\ref{eq:H_R_condition}, the condition for fragmentation becomes:

\begin{equation}
  \frac{H}{R} \lesssim \frac{\pi \Sigma R^2}{M_{\star}} = \frac{\pi \Sigma_o R_o^{p}}{G M_{\star}}R^{(2-p)}.
  \label{eq:analytics_ii}
\end{equation}
Equations~\ref{eq:analytics_i} and~\ref{eq:analytics_ii} show the following:

(i) Increasing the disc mass or decreasing the star mass is likely to promote fragmentation since a greater portion of the disc is likely to fulfil the above criteria.

(ii) The surface mass density profile may play a part in the fragmentation of a disc.  If $p=2$, the disc is scale-free and each radius is equivalent to any other radius: the right hand side (RHS) of equation~\ref{eq:analytics_ii} is constant with radius and the ratio of the cooling timescale to the orbital timescale, $\beta$, is also a constant. Therefore, if the disc settles into a quasi-steady state where the internal heating due to the gravitational instabilities balances the cooling, we expect $Q$ also to be constant with radius (i.e. the left hand side, LHS, of equation~\ref{eq:analytics_ii}, $H/R$, is also a constant).  Consequently, if $p=2$, either the entire disc should settle into a quasi-steady state or the entire disc should fragment.  We note that in a fragmenting disc, because the heating, cooling and fragmentation timescales are all proportional to the dynamical timescale in the disc, the fragmentation should occur first (in absolute terms) at small radii.

For $p < 2$, the RHS of equation~\ref{eq:analytics_ii} increases with radius.  Although, H/R is likely to increase with radius as well, since H/R will typically increase more slowly than the RHS of equation~\ref{eq:analytics_ii}, this condition is more likely to be satisfied in the outer regions of a disc.  Conversely, for $p > 2$, the RHS of equation~\ref{eq:analytics_ii} decreases with radius and hence the condition is more likely to be satisfied at small disc radii.

(iii) For a $p<2$ disc with a low enough $\beta$ such that it can fragment, the exact value of $\beta$ may determine how much of the disc satisfies condition~\ref{eq:analytics_ii}.  If the cooling in a disc is fast such that $\beta$ is small, the temperature and hence sound speed, $c_{\rm s}$ will decrease more rapidly than in a disc where $\beta$ is higher.  Consequently, since $H \propto c_{\rm s}$, the aspect ratio will be lower at any particular radius and hence the disc is more likely to satisfy condition~\ref{eq:analytics_ii} for smaller $\beta$.  Since gravitational instability typically develops on a dynamical timescale, $t_{\rm dyn} \propto 1/\Omega \propto R^{3/2}$, the instability will develop in the inner regions first and therefore fragmentation will first occur as close to the inner regions as possible where the fragmentation criteria are satisfied.  Since more of the disc satisfies the above criteria for decreasing $\beta$, the radius at which the first fragment forms will also decrease.  We therefore expect the fragmentation radius to move inwards with more efficient cooling.

(iv) Crucially, equation~\ref{eq:analytics_ii} shows that the radius is important and that for a shallow surface mass density profile ($p<2$), there does not appear to be a limit for fragmentation (if the disc cools fast enough): provided the disc is large enough, condition~\ref{eq:analytics_ii} will be satisfied (since typically, an increase in $H/R$ with radius will be much smaller than the increase in the RHS of equation~\ref{eq:analytics_ii}).

\section{Numerical setup}
\label{sec:numerics}
Our simulations are carried out using an {\sc sph} code originally developed by \cite{Benz1990} and further developed by \cite*{Bate_Bonnell_Price_sink_ptcls} and \cite{Price_Bate_MHD_h}.  It is a Lagrangian hydrodynamics code, ideal for simulations that require a large range of densities to be followed, such as fragmentation scenarios.

We include the heating effects in the disc due to work done on the gas and artificial viscosity to capture shocks.  The cooling in the disc is taken into account using the cooling parameter, $\beta$ (equation~\ref{eq:beta}), which cools the gas on a timescale given by equation~\ref{eq:tcool}.

In order to model shocks, {\sc sph} requires artificial viscosity.  We use a common form of artificial viscosity by \cite{Monaghan_Gingold_art_vis}, which uses the parameters $\alpha_{\rm SPH}$ and $\beta_{\rm SPH}$.  A corollary of including artificial viscosity is that it adds shear viscosity and causes dissipation.  If this viscosity is too large, the evolution of the disc may be driven artificially, while if it is too small, it will lead to inaccurate modelling of shocks \citep{Matthews_thesis}.  As discussed in \cite{Meru_Bate_opacity}, various values of the {\sc sph} parameters have been tested and we find that a value of $\alpha_{\rm SPH} \sim 0.1$ provides a good compromise between these factors.  Since typically, $\beta_{\rm SPH} \approx 2\alpha_{\rm SPH}$, we choose $\alpha_{\rm SPH}$ and $\beta_{\rm SPH}$ to be 0.1 and 0.2, respectively, which are fixed throughout the simulations.  Furthermore, our work will begin with a comparison with \cite{Rice_beta_condition} and so we use the same values used by them.  We use an adiabatic equation of state with ratio of specific heats, $\gamma = 5/3$.

\subsection{Numerical effects on fragmentation results}
\label{sec:methods_numerics}
\cite{Rice_beta_condition} showed that for a disc with ratio of specific heats, $\gamma = 5/3$, the critical value of the cooling timescale in units of the orbital timescale required for fragmentation is $\beta_{\rm crit} \sim 6-7$.  The {\sc sph} code used for the simulations presented in this paper differs in the way the smoothing length of the particles is set from that of \cite{Rice_beta_condition}: whilst their code sets the smoothing length by approximately fixing the number of neighbours that each particle has to $\approx 50$, the current version uses a variable smoothing length which does not fix the number of neighbours but allows the smoothing length to be spatially adaptive whilst maintaining energy and entropy conservation \citep{Gradh_Springel_Hernquist,Gradh_Monaghan}, with our particular implementation described by \cite{Price_Bate_MHD_h}.

All explicit hydrodynamical simulations must limit the timestep based on the Courant condition.  Our {\sc sph} code also applies a force condition and a viscous timestep condition (see \citealp{SPH_Monaghan} for a review).  In the simulations presented here, since we apply an explicit cooling rate, it is important to ensure that the following timestep criteria is also met:

\begin{equation}
  \Delta t \leq C \frac{\beta}{\Omega},
  \label{eq:tcool_timestep}
\end{equation}
where $C$ is a constant less than unity.  We investigate the effects of varying the constant $C$ on the critical cooling timescale $\beta_{\rm crit}$ by testing values of $C=0.3$, 0.03 and 0.003.  We find that this does not have a significant effect on the fragmentation results and so we use $C=0.3$ for the simulations presented here.  However, the timestep criterion may become more important for small $\beta$ or for particles at small radii.  Therefore, for those simulations carried out with small values of $\beta$ ($\le 3$) or where fragmentation occurs at small radii ($\lesssim 5~AU$), the simulations have been repeated with $C=0.03$ to confirm that this does not play a part in the results.

\section{Benchmarking simulations}
\label{sec:benchmark}

\begin{table*}
  {\footnotesize
  \centering
  \begin{tabular}{llllllll}
    \hline
    Simulation & $\beta$ & $p$ & $q$ & $Q_{\rm min}$ & Initial $Q$ profile & Fragments? \\
    name &  &  & &\\
    \hline
    \hline
    Benchmark1 & 6 & 1 & 0.5 & 2 & Decreasing $Q$ & n\\
    Benchmark2 & 5.5 & 1 & 0.5 & 2 & Decreasing $Q$ & y\\
    Benchmark3 & 5.6 & 1 & 0.5 & 2 & Decreasing $Q$ & n\\
    Benchmark4 & 5 & 1 & 0.5 & 2 & Decreasing $Q$ & y\\
    Benchmark5 & 5 & 1 & -1 & 2 & Flat $Q$ & y\\
    Benchmark6 & 6 & 1 & -1 & 2 & Flat $Q$ & n\\
    Benchmark7 & 5 & 1 & -1 & 1 & Flat $Q$ & y\\
    \hline
  \end{tabular}
  }
  \caption{Summary of the benchmarking simulations described in Section~\ref{sec:benchmark}.  $p$ and $q$ are the initial surface mass density and temperature profiles, $\Sigma \propto R^{-p}$ and $T \propto R^{-q}$, respectively.  Simulations Benchmark1-4 have been set up in the same way as \citet{Rice_beta_condition} whereas simulations Benchmark5-7 have been set up with a uniform Toomre stability profile over the entire disc.}
  \label{tab:benchmark_sim}
\end{table*}

Table~\ref{tab:benchmark_sim} summarises the parameters and fragmentation results of the simulations presented here.  Each simulation was run either beyond the point at which the disc attained a steady state (for $> 6$ outer rotation periods, ORPs), or until it fragmented.

The simulations presented by \cite{Rice_beta_condition} also used an {\sc sph} code.  However, since the way the smoothing length is set in our code differs to the code used by \cite{Rice_beta_condition}, and since it is uncertain as to whether their timestepping considered the cooling timescale, we simulate the same disc that \cite{Rice_beta_condition} simulated in order to initially find the critical cooling timescale in units of the orbital timescale, $\beta_{\rm crit}$.  This is done by setting up a 1~$\rm{M_\odot}$ star with a 0.1~$\rm{M_\odot}$ disc made of 250,000 {\sc sph} particles, spanning $0.25 \le R \le 25$AU.  The initial surface mass density and temperature profiles of the disc are $\Sigma \propto R^{-1}$ and $T \propto R^{-\frac{1}{2}}$, respectively.  The magnitudes of these are set such that the Toomre stability parameter (equation~\ref{eq:Toomre}) at the outer edge of the disc, $Q_{\rm min} = 2$.  This gives an aspect ratio, $H/R \approx 0.05$.  We model the $1~\rm{M_\odot}$ star in the centre of the disc using a sink particle \citep{Bate_Bonnell_Price_sink_ptcls}.  At the inner disc boundary, particles are accreted onto the star if they move within a radius of 0.025~AU of the star or if they move into $0.025 \le R < 0.25 \rm{AU}$ and are gravitationally bound to the star.  At the outer edge, the disc is free to expand.

The simulation was run using a ratio of specific heats, $\gamma = 5/3$ and hence according to \cite{Rice_beta_condition}, $\beta_{\rm crit} \approx 6-7$.  We find that the critical value is $\approx 5.6$ since this is the lowest value of $\beta$ that the discs can have without fragmenting (compare simulations Benchmark1-3).  According to equation~\ref{eq:beta_alpha}, this is equivalent to a critical value of the gravitational stress, $\alpha_{\rm GI,max} \sim 0.07$ which is similar to the value of $\sim 0.06$ obtained by \cite{Rice_beta_condition}.  Given the differences between the codes, we consider this level of agreement acceptable.  We therefore compare our remaining simulations to this value of $\beta_{\rm crit}$.

\begin{figure}
  \includegraphics[width=1.0\columnwidth]{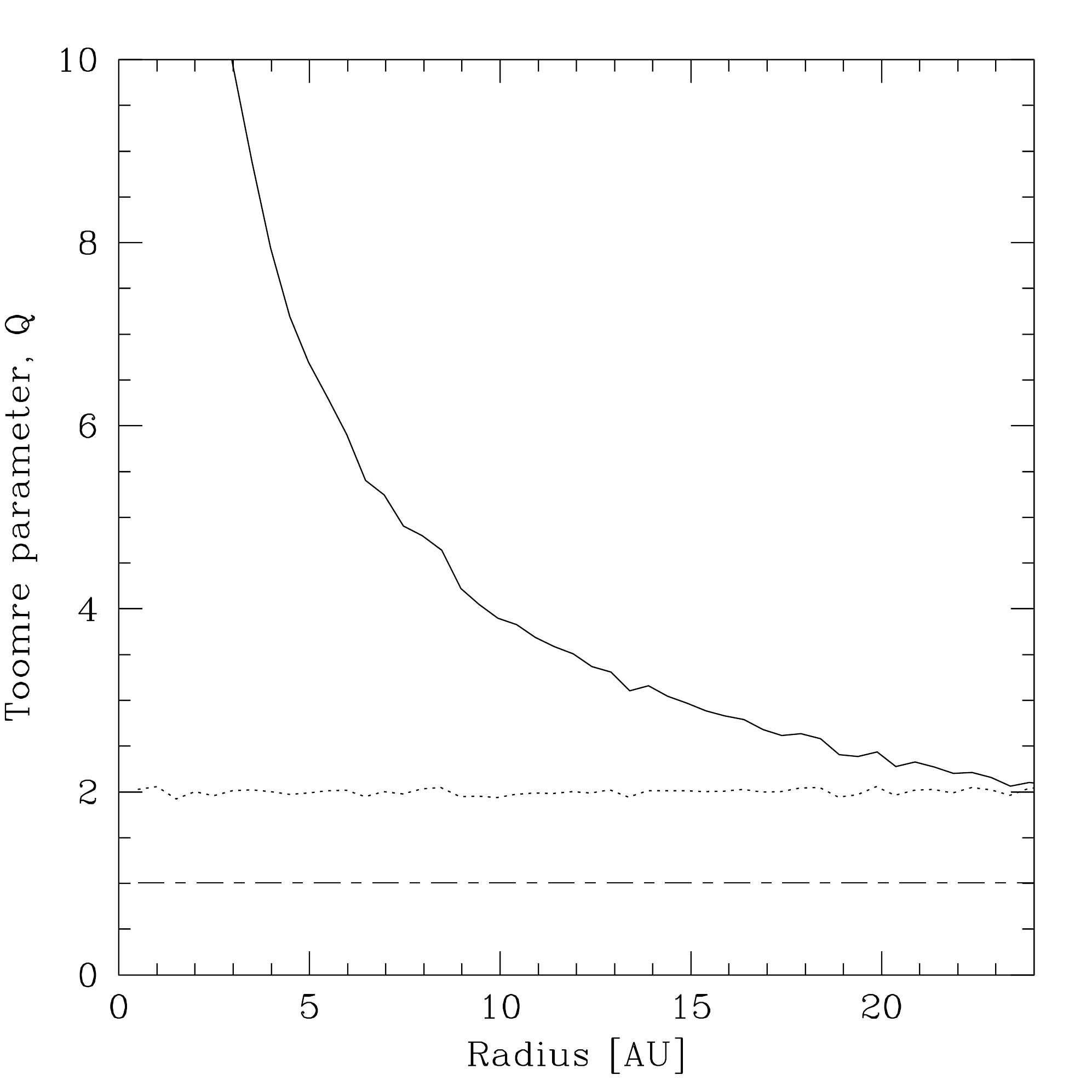}
  \caption{Azimuthally averaged values of the Toomre parameter for the initial discs with decreasing Toomre stability profile (simulations Benchmark1-4), set up in the same way as \citet[][solid line]{Rice_beta_condition}, and with a flat $Q$ profile with $Q=2$ (simulations Benchmark5-7; dotted line).  The critical value of $Q_{\rm crit} = 1$ is also marked.}
  \label{fig:Q_init}
\end{figure}

\begin{figure*}
  \includegraphics[width=2.07\columnwidth]{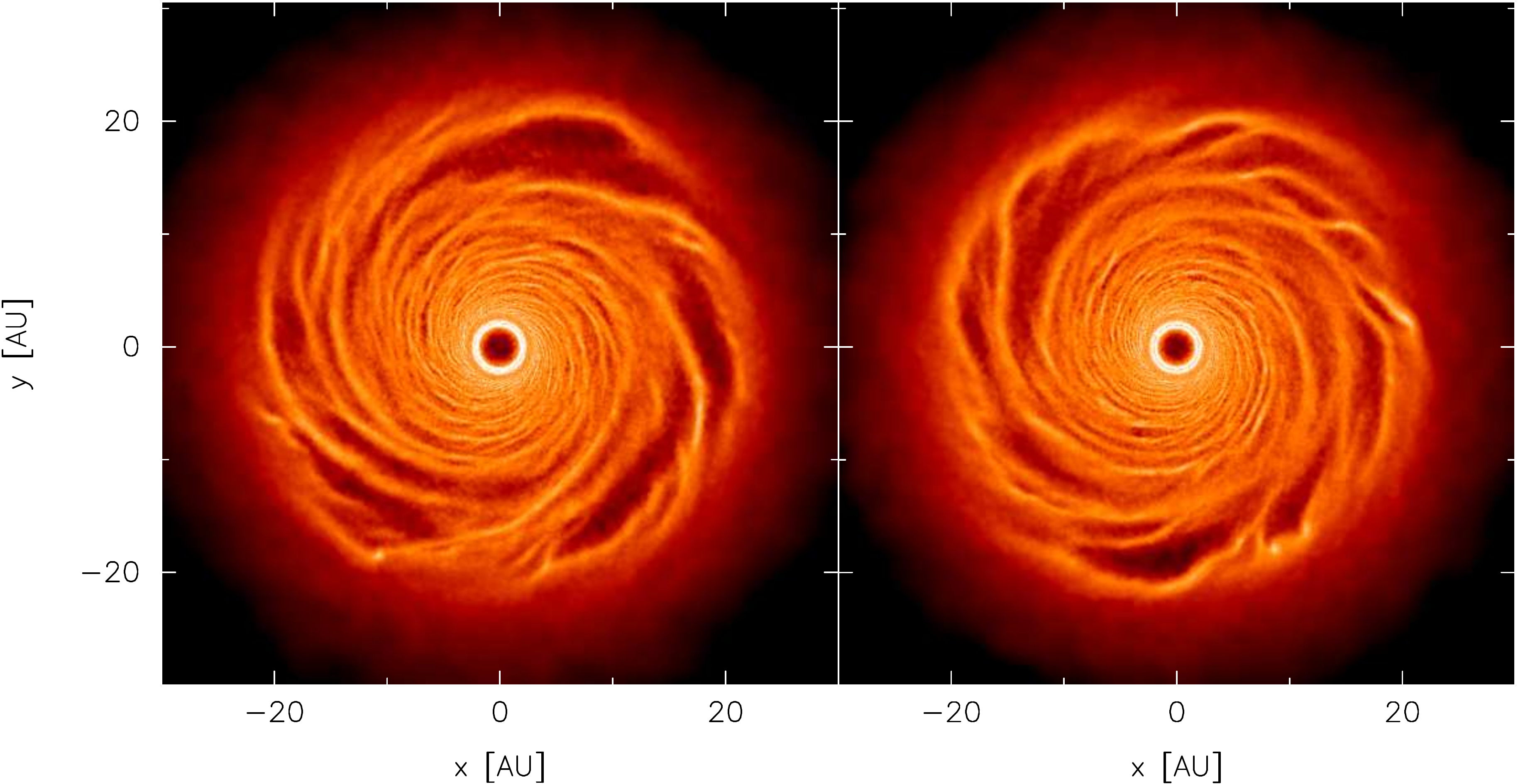}
  \caption{Surface mass density rendered image of the first fragments forming in the simulations using a decreasing Toomre stability profile (simulation Benchmark4, left image) and a disc set up with a flat $Q$ profile with $Q=2$ (simulation Benchmark5, right image).  The discs were run with $\beta=5$.  In both cases the discs first fragment at $R_{\rm f} \approx 20$~AU confirming that the initial temperature profile does not play a part in the evolution of the discs.  The colour scale is a logarithmic scale ranging from log $\Sigma = -7$ (dark) to $-3$ (light) $\rm M_{\odot}/AU^2$.}
  \label{fig:flatQdisc}
\end{figure*}

\begin{figure}
  \includegraphics[width=1.0\columnwidth]{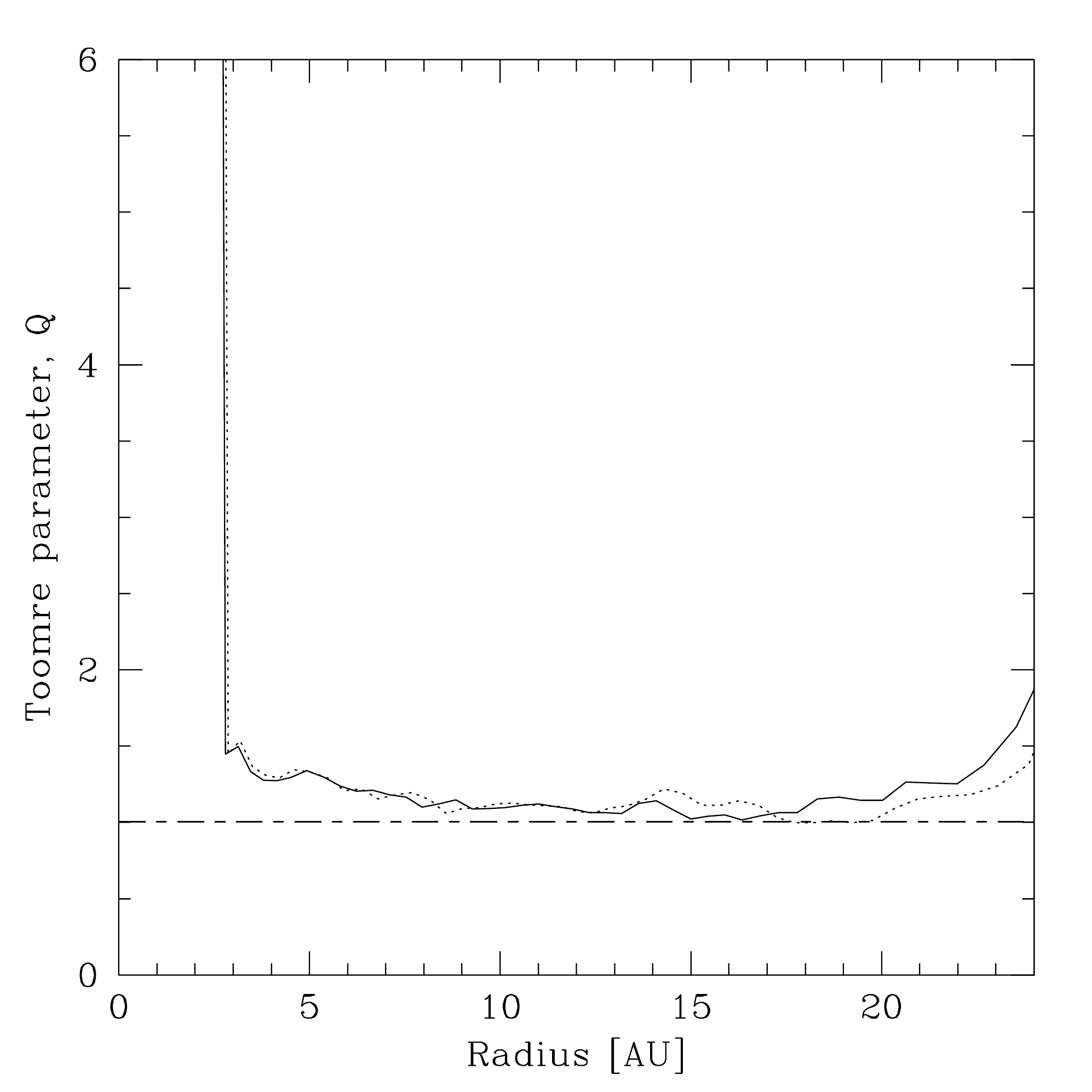}
  \caption{Azimuthally averaged values of the Toomre parameter for the discs with initially decreasing (solid line) and flat (dotted line) Toomre stability profiles (simulations Benchmark1 and 6, respectively).  The discs were run with $\beta=6$.  Despite having different initial temperature profiles, both discs reach a steady-state with very similar Toomre stability profiles, confirming that the initial temperature does not play a part in the evolution of the discs.  The critical value of $Q_{\rm crit} = 1$ is also marked.}
  \label{fig:Q_Retal_flatQ}
\end{figure}

\begin{figure}
  \includegraphics[width=0.9\columnwidth]{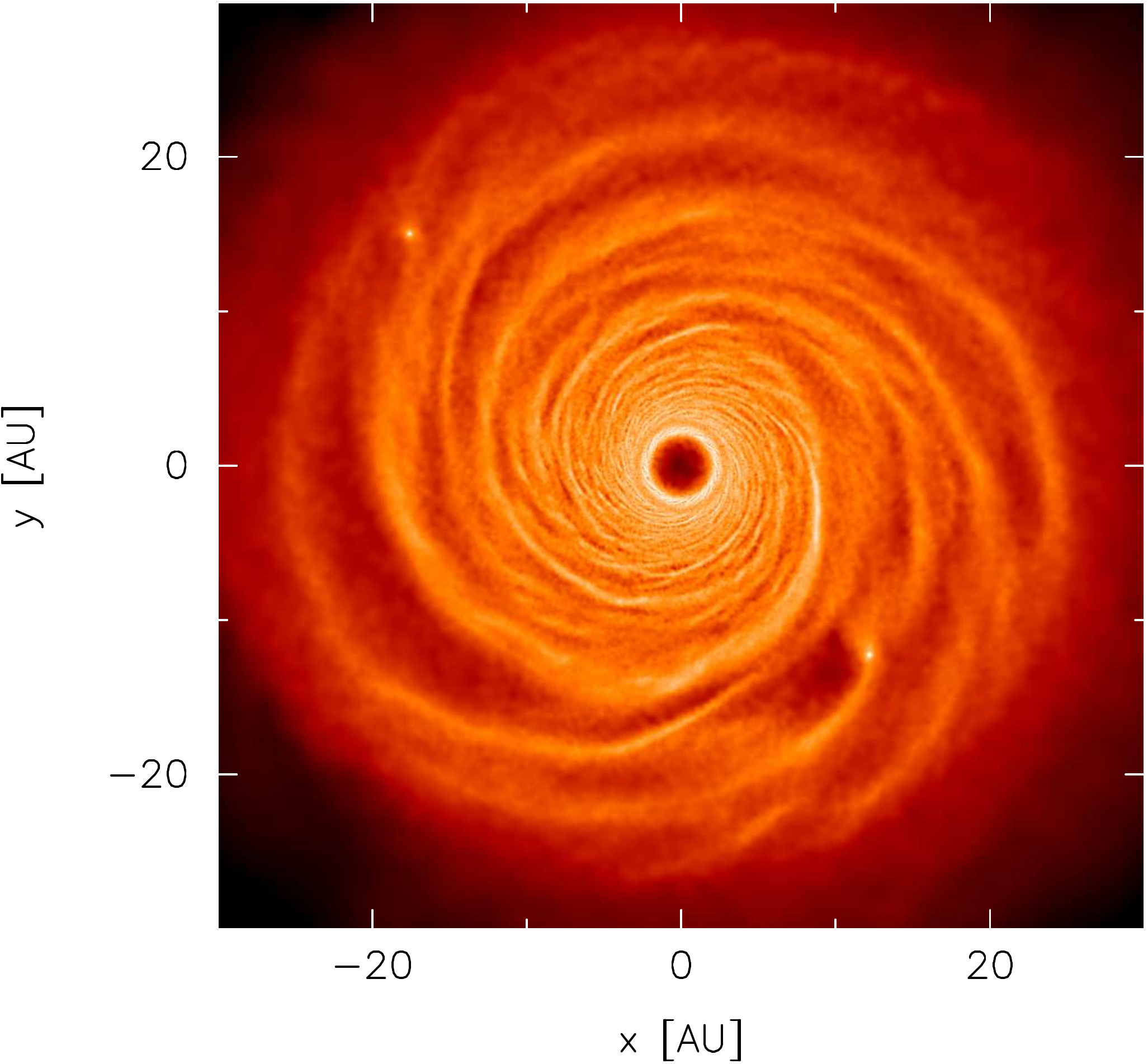}
  \caption{Surface density rendered image of the fragmenting disc in simulation Benchmark7 with an initial flat $Q$ profile with $Q=1$.  The simulation was run with $\beta = 5$.  Despite the initial disc being in a state of marginal stability such that, in theory, any part of the disc may fragment, the disc only fragments in the outer regions.  The colour scale is a logarithmic scale ranging from log $\Sigma = -8$ (dark) to $-3$ (light) $\rm M_{\odot}/AU^2$.}
  \label{fig:FlatQ_Qmin1}
\end{figure}

Figure~\ref{fig:Q_init} shows the initial Toomre stability profile of the \cite{Rice_beta_condition} disc (solid line) that is replicated here.  As a simulation is started, the disc heats up due to the heating from gravitational instability, the resulting compression and viscous heating, and cools on the cooling timescale defined by the cooling parameter, $\beta$.  Consequently, it is expected that the initial disc temperature profile would not play a part in the resulting evolution of the disc.  We therefore test this by setting up a disc with the same surface mass density profile, $\Sigma \propto R^{-1}$, but with a temperature profile, $T \propto R$, so that its initial Toomre stability profile is \emph{flat} (i.e. constant over the entire disc) with $Q=2$ (Figure~\ref{fig:Q_init}; dotted line).  These discs were run for $\beta = 5$ and 6.  Figure~\ref{fig:flatQdisc} shows the images of the evolved disc with decreasing Toomre stability profile and the flat $Q$ disc, run with a cooling time, $\beta = 5$ (simulations Benchmark4 and Benchmark5, respectively).  The first fragments form at $\approx 20$~AU in both discs irrespective of the initial temperature profile.  Figure~\ref{fig:Q_Retal_flatQ} shows the final Toomre stability profiles of both discs run with $\beta = 6$ (simulations Benchmark1 and Benchmark6).  Neither of these discs fragment and both discs evolve into a steady-state with very similar Toomre stability profiles.  It can be seen that the change in initial temperature profile does not make a difference to the final results since with $\beta = 5$ both discs fragment at the same radius and in the non-fragmenting cases, the final Toomre stability profiles are very similar.  Since the temperature in a disc evolves, it is reassuring that the initial temperature profile of the disc does not play a part in the outcome.

As mentioned in Section~\ref{sec:intro}, current wisdom is that according to fragmentation theory, if the Toomre stability parameter is below unity and the timescale on which the disc cools is faster than a critical value, then the disc should fragment.  Therefore, if a disc was set up so that its initial Toomre stability profile was flat with $Q=1$, it would be expected that fragments would form everywhere in the disc soon after the simulation is started.  Figure~\ref{fig:FlatQ_Qmin1} shows the results of this simulation (Benchmark7).  It can be seen that despite starting the simulation in a marginally stable state where any part of the disc may fragment soon after the simulation is started, the disc only fragments in the outer regions.  This illustrates that the disc fragmenting in the outer regions cannot be related to the initial value of the Toomre stability profile, $Q$, and more importantly, fragmentation cannot be a function of $\beta$ alone.

\section{Main simulations}
\label{sec:main_sim}

In this section, we describe the initial conditions for all the individual numerical simulations we have performed to test our analytical predictions from Section~\ref{sec:analytics}.  Table~\ref{tab:main_sim} provides a summary of the initial conditions as well as the radius at which the first fragment forms in the discs that do fragment.

We set up a series of \emph{Reference} discs with $M_{\rm disc} = 0.1\rm M_{\odot}$ consisting of 250,000 {\sc sph} particles, spanning $0.25 \le R \le 25$AU, surrounding a $1\rm M_{\odot}$ star, modelled using a sink particle.   The inner and outer radial disc boundaries have been set up in the same way as the benchmarking discs in Section~\ref{sec:benchmark}.  All the discs in this section have been set up with a flat $Q$ profile.  Therefore, as the surface mass density profile is varied, the initial temperature profile, $q$, is also varied accordingly, where $T \propto R^{-q}$.  The Reference discs have been set up so that $\Sigma \propto R^{-1}$ and $T \propto R$, normalised so that $Q=2$.  We highlight where we have differed from these initial conditions in the remaining simulations.

\begin{table*}
  {\footnotesize
  \centering
  \begin{tabular}{lllllllllll}
    \hline
    Simulation & $\beta$ & $p$ & $\Sigma_o$ & $M_{\rm disc}$ & $M_{\star}$ & $M_{\rm disc}/M_{\star}$ & $Q$ & Disc & $R_{\rm f}$ & $\frac{\Sigma R_{\rm f}^2}{M_{\star}}$\\
   name & & & [$\rm M_{\odot}/(AU)^2$] & & & & & radius [AU] & [AU] &\\
   \hline
   \hline
   Reference-beta6 & 6 & 1 & $6.4 \times 10^{-4}$ & 0.1 & 1 & 0.1 & 2 & 25 & - & - \\
    Reference-beta5.5 & 5.5 & 1 & $6.4 \times 10^{-4}$ & 0.1 & 1 & 0.1 & 2 & 25 & 22 & $1.4 \times 10^{-2}$\\
    Reference-beta5 & 5 & 1 & $6.4 \times 10^{-4}$ & 0.1 & 1 & 0.1 & 2 & 25 & 20 & $1.3 \times 10^{-2}$\\
    Reference-beta4 & 4 & 1 & $6.4 \times 10^{-4}$ & 0.1 & 1 & 0.1 & 2 & 25 & 20 & $1.3 \times 10^{-2}$\\
    Reference-beta3 & 3 & 1 & $6.4 \times 10^{-4}$ & 0.1 & 1 & 0.1 & 2 & 25 & 8 & $5.1 \times 10^{-3}$\\
    Reference-beta2 & 2 & 1 & $6.4 \times 10^{-4}$ & 0.1 & 1 & 0.1 & 2 & 25 & 3 & $1.9 \times 10^{-3}$\\
    Reference-beta1 & 1 & 1 & $6.4 \times 10^{-4}$ & 0.1 & 1 & 0.1 & 2 & 25 & 2.5 & $1.6 \times 10^{-3}$\\
    \hline
    p1.5-beta4 & 4 & 1.5 & $1.8 \times 10^{-3}$ & 0.1 & 1 & 0.1 & 2 & 25 & - & -\\
    p1.5-beta3.5 & 3.5 & 1.5 & $1.8 \times 10^{-3}$ & 0.1 & 1 & 0.1 & 2 & 25 & 18 & $7.5 \times 10^{-3}$\\
    p1.5-beta3 & 3 & 1.5 & $1.8 \times 10^{-3}$ & 0.1 & 1 & 0.1 & 2 & 25 & 1.7 & $2.3 \times 10^{-3}$\\
    p2-beta3.5 & 3.5 & 2 & $3.5 \times 10^{-3}$ & 0.1 & 1 & 0.1 & 2 & 25 & - & - \\
    p2-beta3 & 3 & 2 & $3.5 \times 10^{-3}$ & 0.1 & 1 & 0.1 & 2 & 25 & 0.45 & $3.5 \times 10^{-3}$\\
    p2-beta2 & 2 & 2 & $3.5 \times 10^{-3}$ & 0.1 & 1 & 0.1 & 2 & 25 & 0.3 & $3.5 \times 10^{-3}$ \\
    p2.5-beta4 & 4 & 2.5 & $4.4 \times 10^{-3}$ & 0.1 & 1 & 0.1 & 2 & 25 & - & - \\
    p2.5-beta3.5 & 3.5 & 2.5 & $4.4 \times 10^{-3}$ & 0.1 & 1 & 0.1 & 2 & 25 & 0.4 & $7.0 \times 10^{-3}$\\
    p2.5-beta3.5-Q5 & 3.5 & 2.5 & $4.4 \times 10^{-3}$ & 0.1 & 1 & 0.1 & 5 & 25 & 0.3 & $8.1 \times 10^{-3}$\\
    p2.5-beta3 & 3 & 2.5 & $4.4 \times 10^{-3}$ & 0.1 & 1 & 0.1 & 2 & 25 & 0.3 & $8.1 \times 10^{-3}$\\
    p2.5-beta2 & 2 & 2.5 & $4.4 \times 10^{-3}$ & 0.1 & 1 & 0.1 & 2 & 25 & 0.35 & $7.5 \times 10^{-3}$\\
    \hline
    p1-Mstar2 & 5 & 1 & $6.4 \times 10^{-4}$ & 0.1 & 2 & 0.05 & 2 & 25 & - & -\\
    p1-Mstar0.5 & 5 & 1 & $6.4 \times 10^{-4}$ & 0.1 & 0.5 & 0.2 & 2 & 25 & 13 & $1.7 \times 10^{-2}$\\
    \hline
    p1-Mdisc0.2 & 5 & 1 & $1.3 \times 10^{-3}$ & 0.2 & 1 & 0.2 & 2 & 25 & 14 & $1.8 \times 10^{-2}$\\
    p1-Mdisc0.05 & 5 & 1 & $3.2 \times 10^{-4}$ & 0.05 & 1 & 0.05 & 2 & 25 & - & -\\
   p1-beta1-Mdisc0.01 & 1 & 1 & $6.4 \times 10^{-5}$ & 0.01 & 1 & 0.01 & 2 & 25 & 6.5 & $4.2 \times 10^{-4}$\\
    p1-beta2-Mdisc0.01 & 2 & 1 & $6.4 \times 10^{-5}$ & 0.01 & 1 & 0.01 & 2 & 25 & 15 & $9.6 \times 10^{-4}$\\
    p1-beta2.5-Mdisc0.01 & 2.5 & 1 & $6.4 \times 10^{-5}$ & 0.01 & 1 & 0.01 & 2 & 25 & 17 & $1.1 \times 10^{-3}$\\
    p1-beta3-Mdisc0.01 & 3 & 1 & $6.4 \times 10^{-5}$ & 0.01 & 1 & 0.01 & 2 & 25 & - & -\\
    p1-beta8-Mdisc0.3 & 8 & 1 & $1.9 \times 10^{-3}$ & 0.3 & 1 & 0.3 & 2 & 25 & - & -\\
    p1-beta10-Mdisc0.5 & 10 & 1 & $3.2 \times 10^{-3}$ & 0.5 & 1 & 0.5 & 2 & 25 & - & -\\
    p1-beta5-Mdisc1 & 5 & 1 & $6.4 \times 10^{-3}$ & 1 & 1 & 1 & 2 & 25 & 5.5 & $3.5 \times 10^{-2}$\\
    p1-beta7-Mdisc1 & 7 & 1 & $6.4 \times 10^{-3}$ & 1 & 1 & 1 & 2 & 25 & - & -\\
    p1-beta10-Mdisc1 & 10 & 1 & $6.4 \times 10^{-3}$ & 1 & 1 & 1 & 2 & 25 & - & -\\
    p1-beta15-Mdisc1 & 15 & 1 & $6.4 \times 10^{-3}$ & 1 & 1 & 1 & 2 & 25 & - & -\\
    \hline
    p1-beta6-extended & 6 & 1 & $6.4 \times 10^{-4}$ & 0.2 & 1 & 0.2 & 2 & 50 & 24.5 & $1.6 \times 10^{-2}$\\
    p1-beta7-extended & 7 & 1 & $6.4 \times 10^{-4}$ & 0.2 & 1 & 0.2 & 2 & 50 & 29 & $1.9 \times 10^{-2}$\\
    p1-beta8-extended & 8 & 1 & $6.4 \times 10^{-4}$ & 0.2 & 1 & 0.2 & 2 & 50 & 30 & $1.9 \times 10^{-2}$\\
    p1.5-beta4-extended & 4 & 1.5 & $1.8 \times 10^{-3}$ & 0.15 & 1 & 0.15 & 2 & 50 & 33 & $1.0 \times 10^{-2}$\\
    p1-beta6-Mdisc0.1-extended & 6 & 1 & $3.2 \times 10^{-4}$ & 0.1 & 1 & 0.1 & 2 & 50 & 40 & $1.3 \times 10^{-2}$\\
    p1-beta6-Mdisc0.2-Mstar2-extended & 6 & 1 & $6.4 \times 10^{-4}$ & 0.2 & 2 & 0.1 & 2 & 50 & 34 & $1.1 \times 10^{-2}$\\
   \hline
  \end{tabular}
  }
  \caption{Summary of the main simulations.  $p$ describes the initial surface mass density profile, $\Sigma \propto R^{-p}$, and $\Sigma_o$ is the normalisation constant required to produce a disc with mass $M_{\rm disc}$.  The final column represents the RHS of equation~\ref{eq:analytics_ii} for the location at which the first fragment forms, $R_{\rm f}$.  The simulations were run with an initial flat Toomre stability profile, $Q$.}
  \label{tab:main_sim}
\end{table*}

The analytical work presented in Section~\ref{sec:analytics} suggests that for shallow surface mass density profiles, $p<2$, fragments would form in the outer regions of the discs, whilst for discs with steeper surface mass density profiles, $p \gtrsim 2$, the disc would fragment in the inner regions.  We therefore test different values of the slope of the surface mass density profiles using $p=1$ (simulations Reference-beta5.5 and Reference-beta6), 1.5 (simulations p1.5-beta3.5 and p1.5-beta4), 2.0 (simulations p2-beta2, p2-beta3 and p2-beta3.5) and 2.5 (simulations p2.5-beta2, p2.5-beta3.5 and p2.5-beta4).  In addition, we also carry out a further simulation which is the same as simulation p2.5-beta3.5 but with an initial flat $Q$ profile, $Q=5$ (i.e. so that the initial temperature is 25/4 times hotter than the disc in simulation p2.5-beta3.5), to test the effects of an initially hotter disc on the location of fragmentation (simulation p2.5-beta3.5-Q5).

The analysis also suggests that for a disc with a fast enough cooling timescale such that it would fragment, the location at which the first fragment would form would move inwards to smaller radii as the cooling timescale is decreased.  We therefore test the effect of decreasing $\beta$ on the fragment location by simulating the Reference disc (i.e. with a surface mass density profile, $p=1$) with different values of the cooling timescale, $\beta$ = 5.5, 5, 4, 3, 2 and 1 (simulations Reference-beta5.5, Reference-beta5, Reference-beta4, Reference-beta3, Reference-beta2 and Reference-beta1, respectively).

We also argued that varying the disc or star mass would affect fragmentation.  We test the effects of doubling and halving the star mass in simulations p1-Mstar2 and p1-Mstar0.5, respectively, and compare these to the Reference-beta5 simulation.   We also carry out extensive tests of the effects of varying the disc mass firstly by doubling and halving the disc mass (simulations p1-Mdisc0.2 and p1-Mdisc0.05, respectively) and secondly by considering more extreme disc masses of $0.01\rm M_{\odot}$ (simulations p1-beta0.3-Mdisc0.01, p1-beta1-Mdisc0.01, p1-beta2-Mdisc0.01, p1-beta2.5-Mdisc0.01 and p1-beta3-Mdisc0.01), $0.3 \rm M_{\odot}$ (simulation p1-beta8-Mdisc0.3), $0.5 \rm M_{\odot}$ (simulation p1-beta10-Mdisc0.5) and $1\rm M_{\odot}$ (simulations p1-beta5-Mdisc1, p1-beta10-Mdisc1 and p1-beta15-Mdisc1) whilst maintaining a central star mass of $1\rm M_{\odot}$.

\begin{figure}
  \includegraphics[width=1.0\columnwidth]{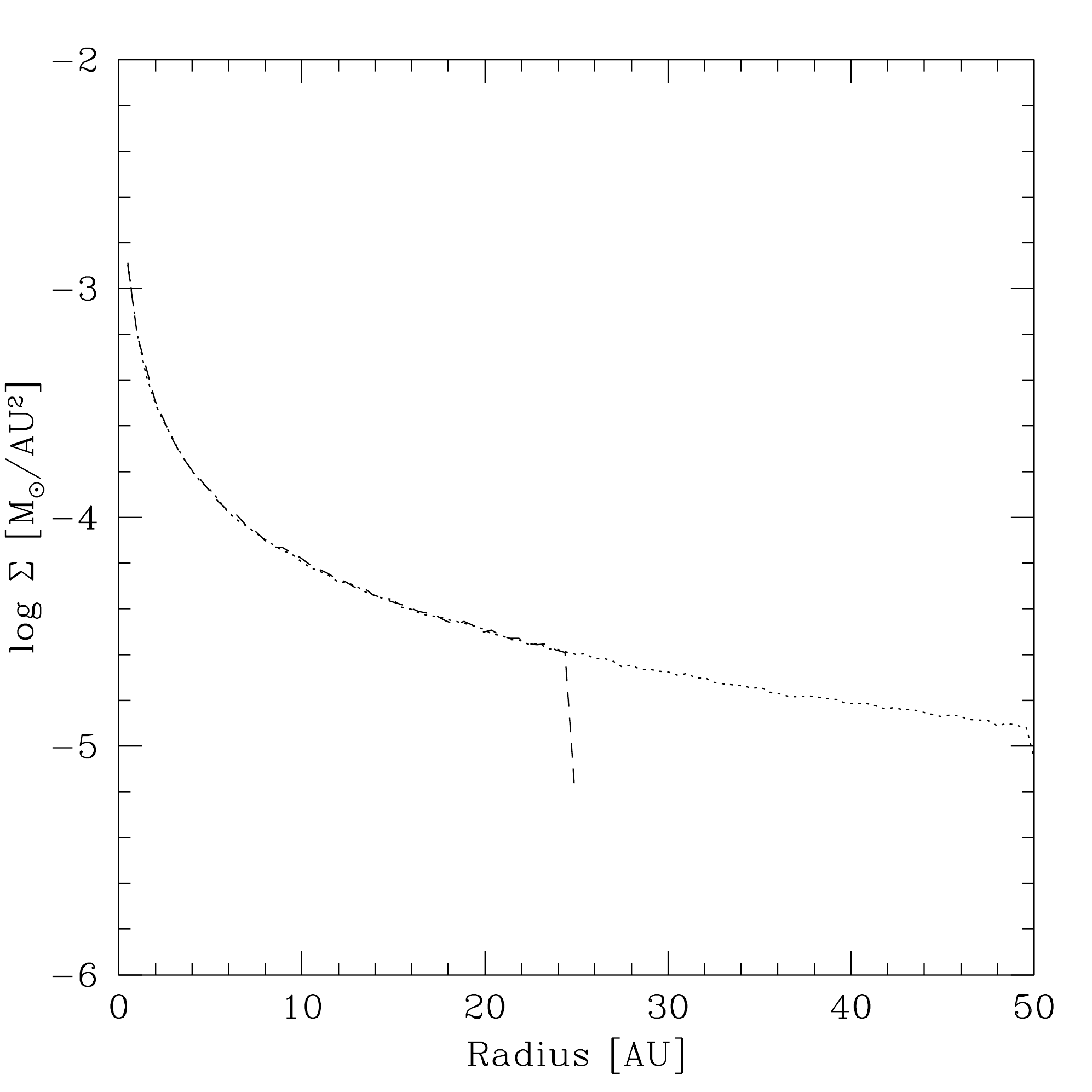}
  \caption{Initial surface mass density profiles of the discs used in simulations p1-beta6 and p1-beta6-extended.  The extended disc has the same surface mass density profile as the smaller disc.}
  \label{fig:sigms_normal_extended}
\end{figure}

The analytical work presented in Section~\ref{sec:analytics} also showed that for a shallow surface mass density profile ($p<2$), the radius of the disc may be the limiting aspect that causes a disc not to fragment.  We therefore test a series of \emph{extended discs} which have outer radii, $R_{\rm out} = 50$AU.  Simulations p1-beta6-extended, p1-beta7-extended and p1-beta8-extended are set up so that $\Sigma_o$ and $p$ are the same as in the Reference discs (Figure~\ref{fig:sigms_normal_extended}).  However, to extend the disc to $R_{\rm out}=50$AU, the disc masses are increased to $M_{\rm disc} = 0.2 \rm M_{\odot}$.  We run this simulation for $\beta = 6$, 7 and 8 (i.e. values that are larger than the critical values identified in Section~\ref{sec:benchmark}).  (Note that in order to keep the mass of the individual {\sc sph} particles the same as in the Reference simulations, we use 500,000 particles to make up this disc).  In addition we also set up an extended disc with a surface mass density profile, $p=1.5$, which has a disc mass of $1.5\rm M_{\odot}$ (so that $\Sigma_o$ is the same as in simulation p1.5-beta4).  We run this simulation using $\beta = 4$ (simulation p1.5-beta4-extended).  (As before, since we wish to keep the mass of the {\sc sph} particles the same as in simulation p1.5-beta4, we use 375,000 particles in this disc.)

Furthermore, we progress the analysis of extended discs by simulating two further discs (using 500,000 particles): the first is the same as that in p1-beta6-extended but using a total disc mass of $M_{\rm disc} = 0.1 \rm M_{\odot}$ (simulation p1-beta6-Mdisc0.1-extended) so that the total disc mass is the same as in p1-beta6 but $\Sigma_o$ is smaller.  The second is also the same as in p1-beta6-extended but the central star mass is also $M_{\star} = 2\rm M_{\odot}$ so that the disc to star mass ratio is kept constant (simulation p1-beta6-Mdisc0.2-Mstar2-extended).  Both of these discs are run with $\beta = 6$.

\section{Results}
\label{sec:results}
For the analysis that follows, the key aspect about the fragments that will be considered will be the first fragment that forms.  This is because subsequent evolution of the disc following an initial fragmentation stage may involve additional complexities that are beyond the scope of this paper.  Table~\ref{tab:main_sim} summarises the key fragmenting results.  The radius at which the first fragments form, $R_{\rm f}$, has been determined by eye from the disc images.  It is important to note that as seen in past simulations (e.g. \citealp{Lodato_Rice_original, Meru_Bate_opacity}), the surface mass density profile does not change significantly during the simulations, particularly for low mass discs ($M_{\rm disc} \lesssim 0.2 \rm M_{\odot}$).  We highlight where the surface mass density profiles do change significantly and discuss the effects of this.

\subsection{Fragmentation dependency on the surface mass density profile}
\label{sec:p}

\begin{figure}
  \centering
  \includegraphics[width=0.98\columnwidth]{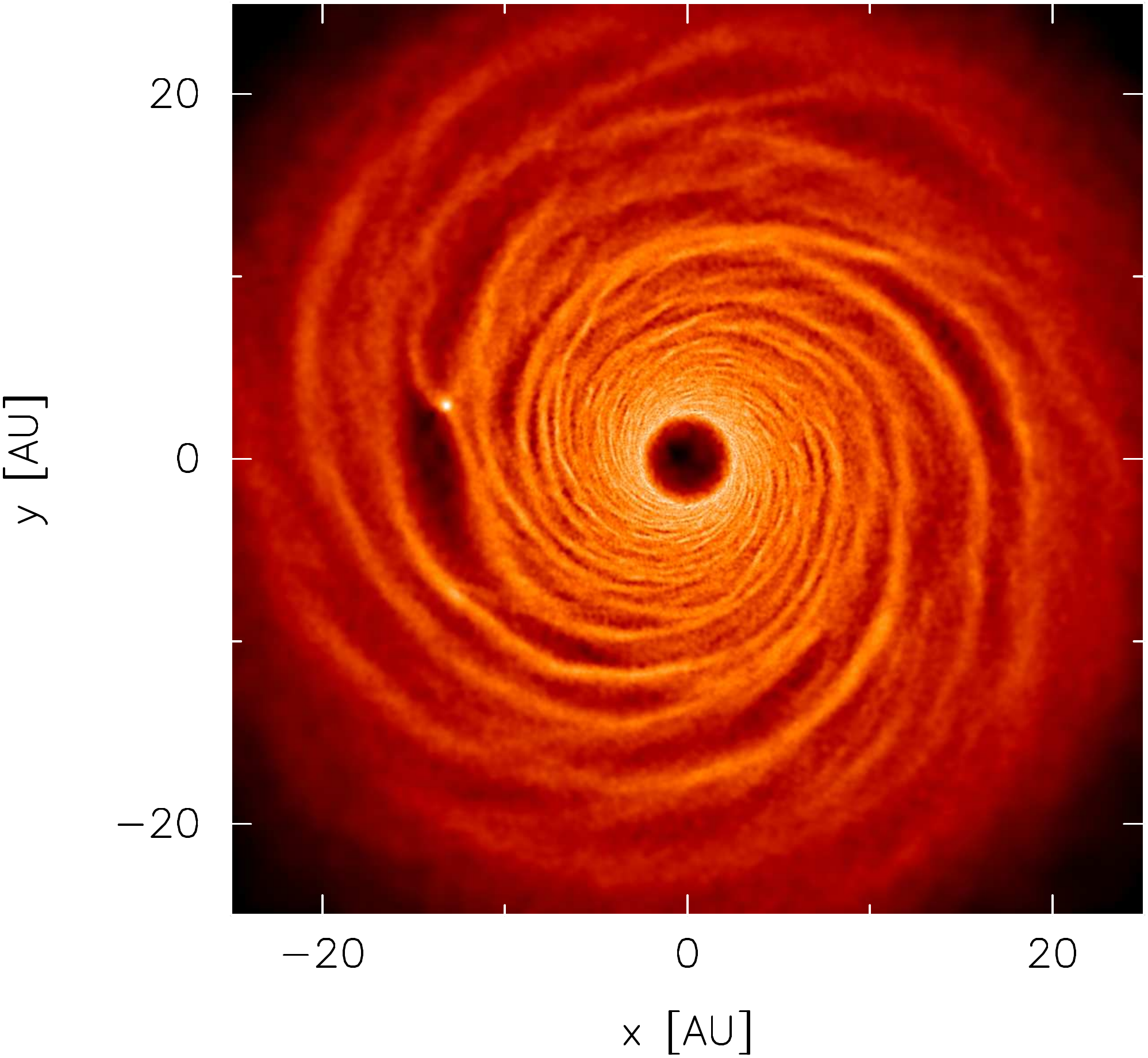}
  \caption{Surface mass density rendered image of the fragmenting disc with initial surface mass density profile $\Sigma \propto R^{-1}$.  The simulation (Reference-beta5.5) used $\beta=5.5$.  The fragment forms in the outer regions of the disc, confirming the analytical predictions in Section~\ref{sec:analytics}.  The colour scale is a logarithmic scale ranging from log $\Sigma = -6$ (dark) to $-3$ (light) $\rm M_{\odot}/AU^2$.}
  \label{fig:S-1}
\end{figure}

\begin{figure}
 \includegraphics[width=0.98\columnwidth]{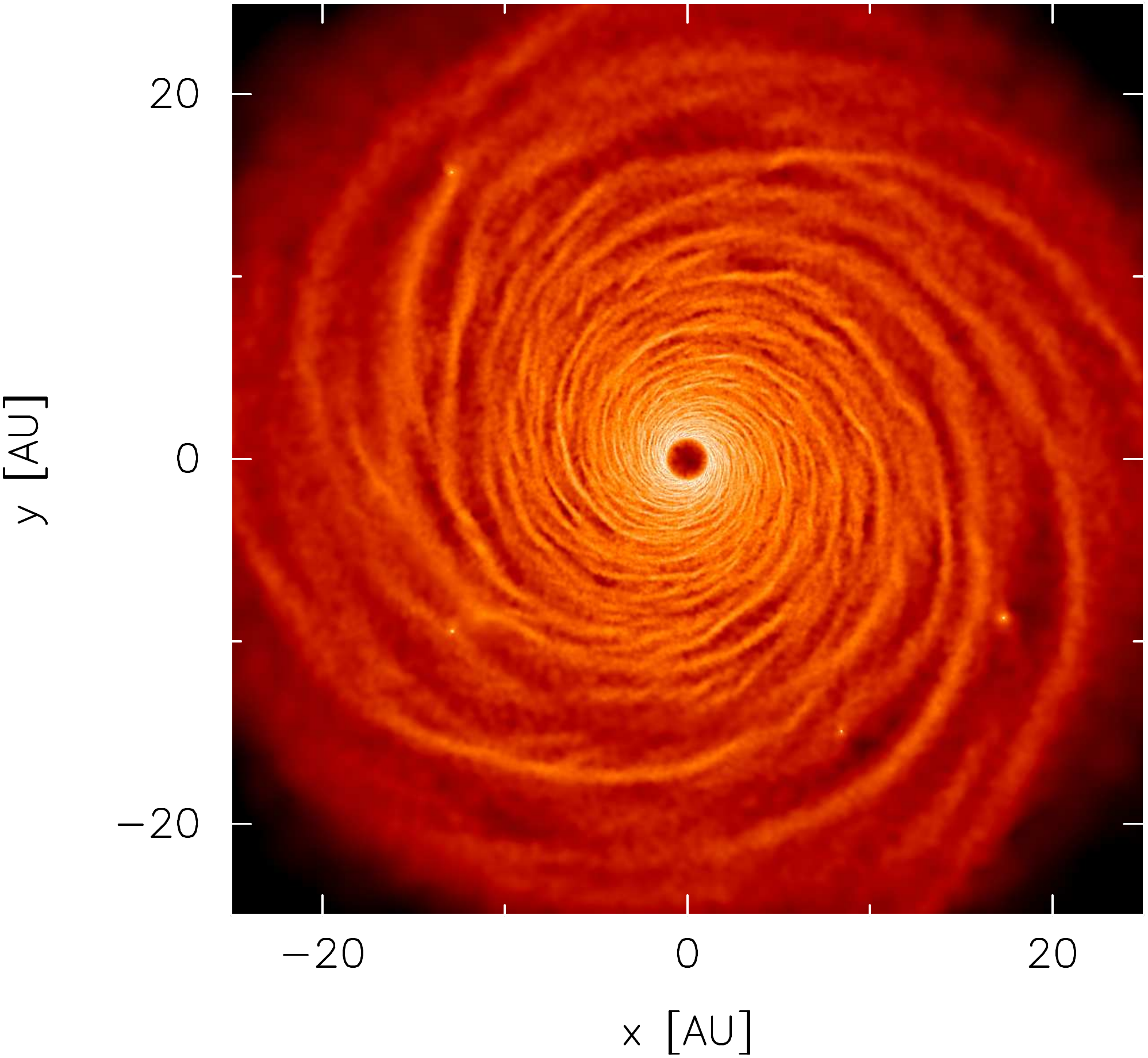}
  \caption{Surface mass density rendered image of the fragmenting disc with initial surface mass density profile $\Sigma \propto R^{-3/2}$.  The simulation used $\beta=3.5$.  The fragment forms in the outer regions of the disc, confirming the analytical predictions in Section~\ref{sec:analytics}.  The colour scale is a logarithmic scale ranging from log $\Sigma = -7$ (dark) to $-2$ (light) $\rm M_{\odot}/AU^2$.}
  \label{fig:S-1.5}
\end{figure}

\begin{figure*}
  \includegraphics[width=2.1\columnwidth]{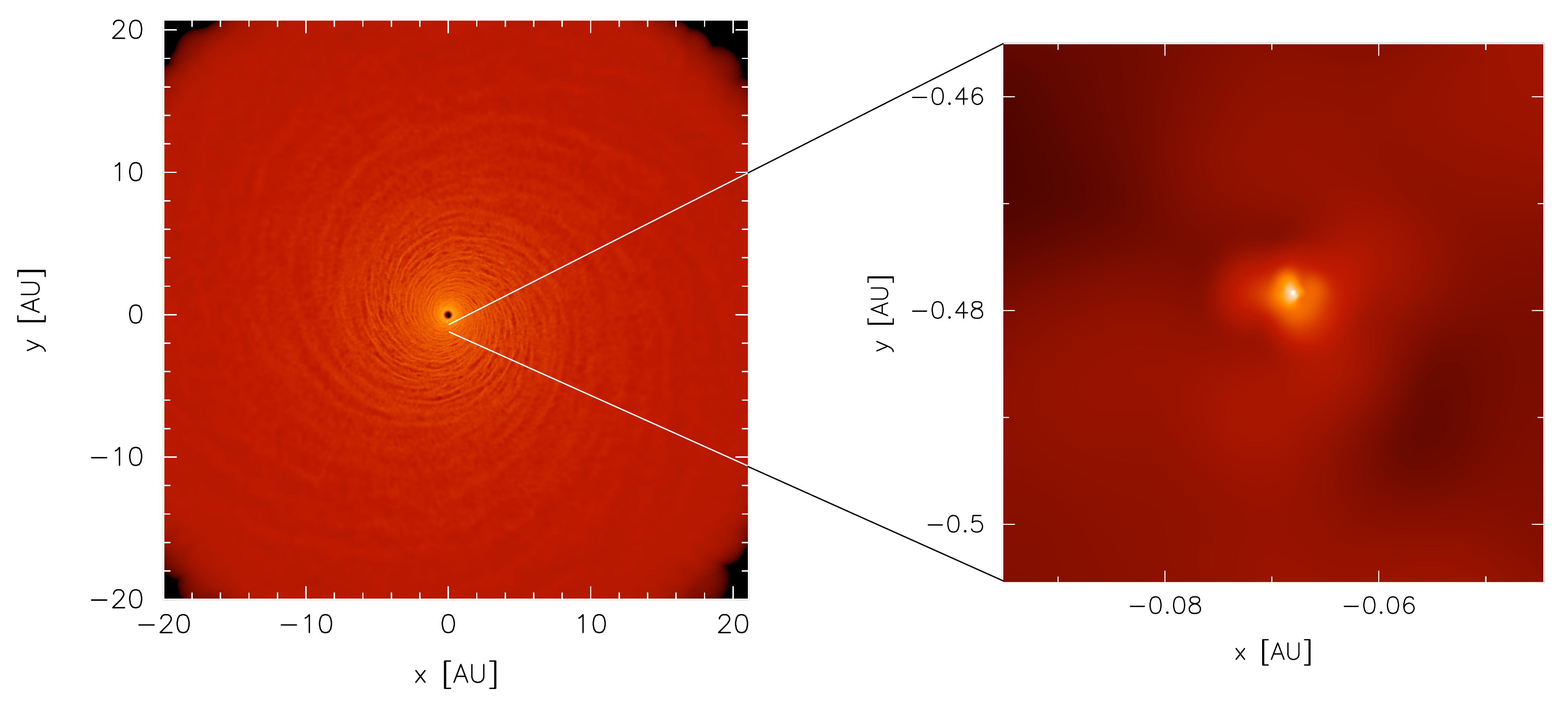}
  \caption{Surface mass density rendered image of the fragmenting disc with initial surface mass density profile $\Sigma \propto R^{-2}$ (simulation p2-beta3).  The simulation used $\beta=3$.  The fragment forms in the inner regions of the disc as shown by the zoomed in image of the disc, confirming the analytical predictions in Section~\ref{sec:analytics}.  The colour scale is a logarithmic scale ranging from log $\Sigma = -11$ (dark) to $2$ (light) $\rm M_{\odot}/AU^2$ in the zoomed out image and from log $\Sigma = -3.5$ (dark) to $1$ (light) $\rm M_{\odot}/AU^2$ in the zoomed in image.}
 \label{fig:S-2}
\end{figure*}

\begin{figure*}
  \includegraphics[width=2.1\columnwidth]{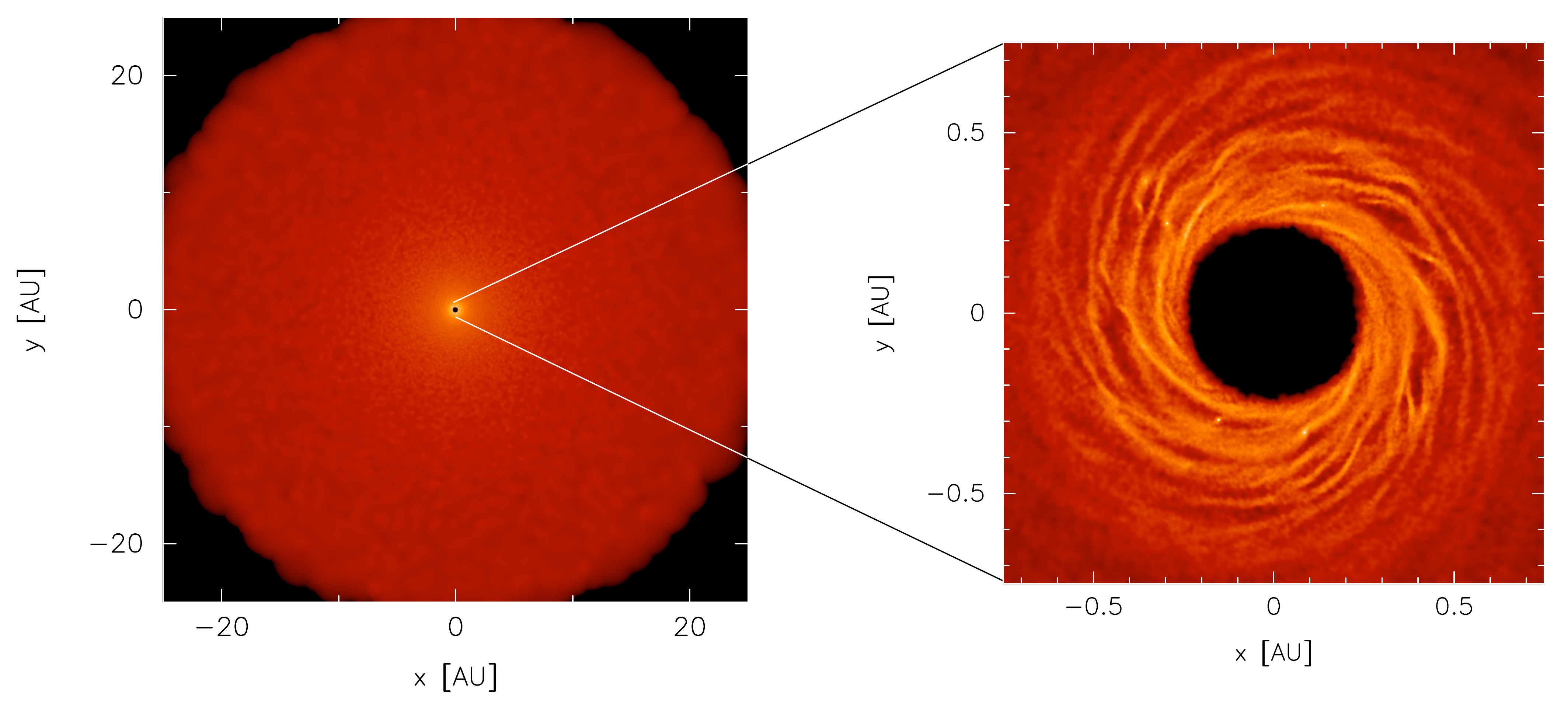}
  \caption{Surface mass density rendered image of the fragmenting disc with initial surface mass density profile $\Sigma \propto R^{-5/2}$ (simulation p2.5-beta3.5).  The simulation used $\beta=3.5$.  The fragment forms in the inner regions of the disc, confirming the analytical predictions in Section~\ref{sec:analytics}.  The colour scale is a logarithmic scale ranging from log $\Sigma = -12$ (dark) to $3$ (light) $\rm M_{\odot}/AU^2$ in the zoomed out image and from log $\Sigma = -4$ (dark) to $0.4$ (light) $\rm M_{\odot}/AU^2$ in the zoomed in image.}
  \label{fig:S-2.5}
\end{figure*}

\begin{figure}
  \includegraphics[width=1.0\columnwidth]{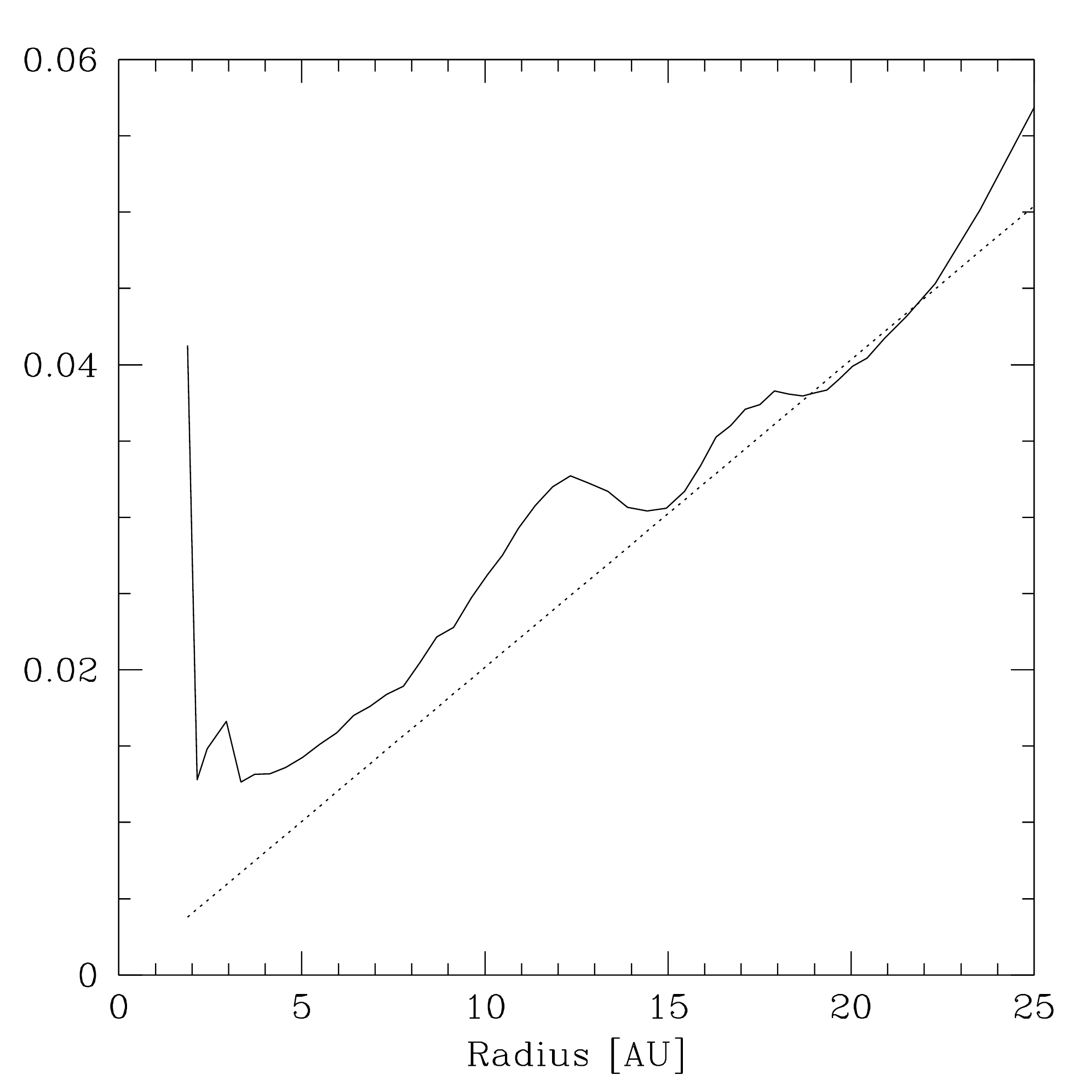}
  \caption{Plot of disc aspect ratio, H/R (solid line), against the RHS of equation~\ref{eq:analytics_ii} (dotted line) for simulation Reference-beta5.5.  Condition~\ref{eq:analytics_ii} is satisfied at $\approx 20$AU where the disc first fragments, confirming the analytical predictions in Section~\ref{sec:analytics}.}
  \label{fig:H_R_S-1}
\end{figure}

\begin{figure}
  \includegraphics[width=1.0\columnwidth]{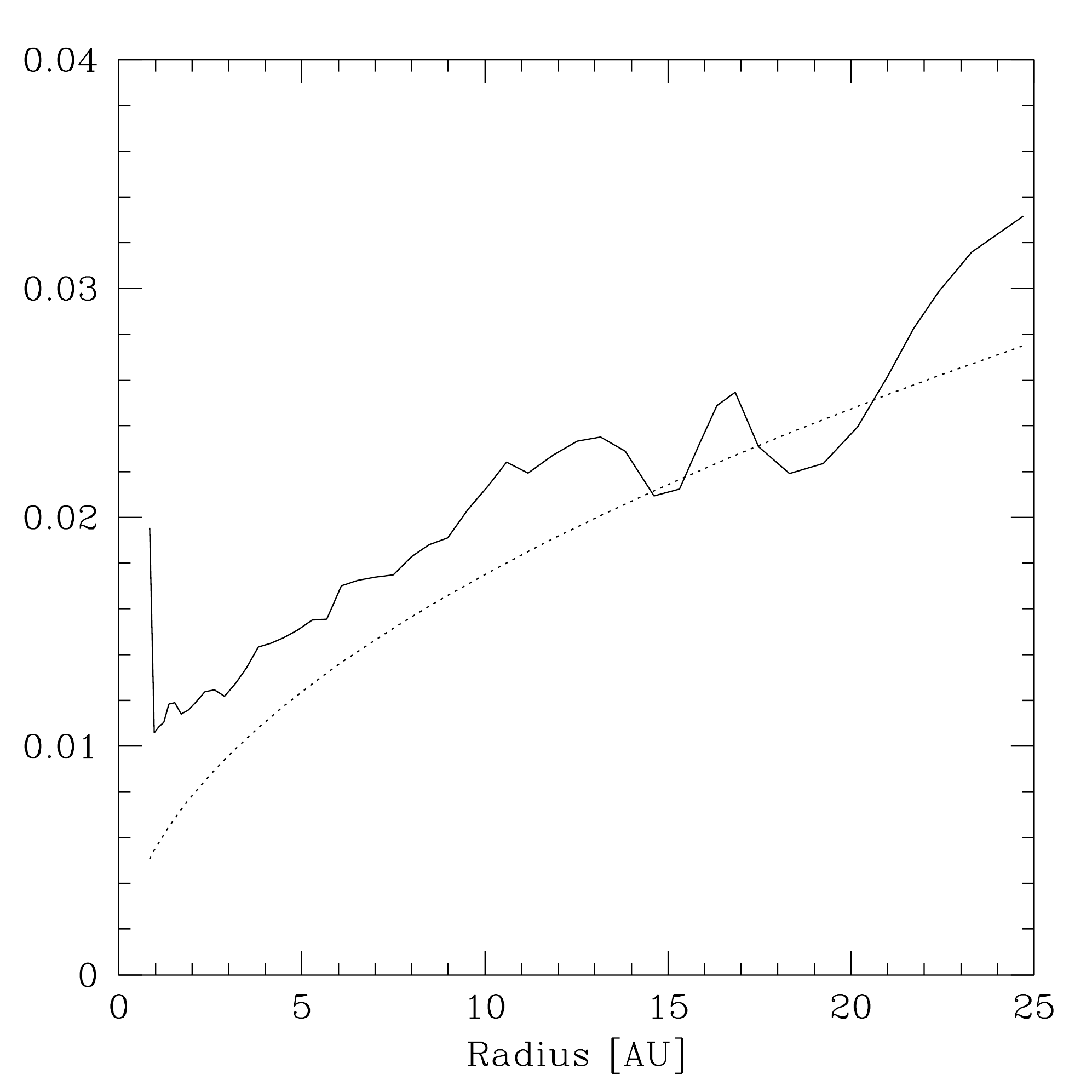}
  \caption{Plot of disc aspect ratio, H/R (solid line), against the RHS of equation~\ref{eq:analytics_ii} (dotted line) for simulation p1.5-beta3.5.  Condition~\ref{eq:analytics_ii} is satisfied at $\approx 19$AU where the disc first fragments, confirming the analytical predictions in Section~\ref{sec:analytics}.}
  \label{fig:H_R_S-1.5}
\end{figure}

Figures~\ref{fig:S-1}-\ref{fig:S-2.5} show that as the surface mass density profile steepens, the location at which the first fragment forms moves to smaller radii in the disc.  The analytical theory presented in Section~\ref{sec:analytics} shows that for a shallow surface mass density fall off where $p<2$, the fragments form in the outer regions of the disc (provided the cooling criterion is also satisfied).  This is indeed the case for simulations with $p=1$ and $p=1.5$ (simulations Reference-beta5.5 and p1.5-beta3.5, respectively) as Figures~\ref{fig:S-1} and~\ref{fig:S-1.5} show that the fragments form at $R_{\rm f}\approx 20$~AU and  $\approx 19$~AU, respectively.  Figures~\ref{fig:H_R_S-1} and~\ref{fig:H_R_S-1.5} show the radial profile of the aspect ratios (calculated by azimuthally averaging the sound speed at each radii and dividing by $\Omega R$ at that radii) in the discs compared to the RHS of equation~\ref{eq:analytics_ii} at a time shortly before the discs fragment.  It can be seen that condition~\ref{eq:analytics_ii} is satisfied at the region in which the first fragment forms shortly after.  The oscillations in $H/R$ are due to temperature fluctuations since although the cooling rate in the disc changes smoothly with radius, the heating of the disc occurs primarily in the spiral shocks.  This therefore confirms the analytical predictions for shallow surface mass density profiles ($p<2$) presented in Section~\ref{sec:analytics}.  It is important to note that for a flat $Q$ profile, the temperature profiles in the discs are an increasing function of radius ($T \propto R$) for $p=1$ and a constant temperature profile for $p=1.5$, yet the discs still fragment in the outer regions (again, re-emphasising that the initial temperature profile does not play a part in the disc evolution).

\begin{figure}
  \includegraphics[width=1.0\columnwidth]{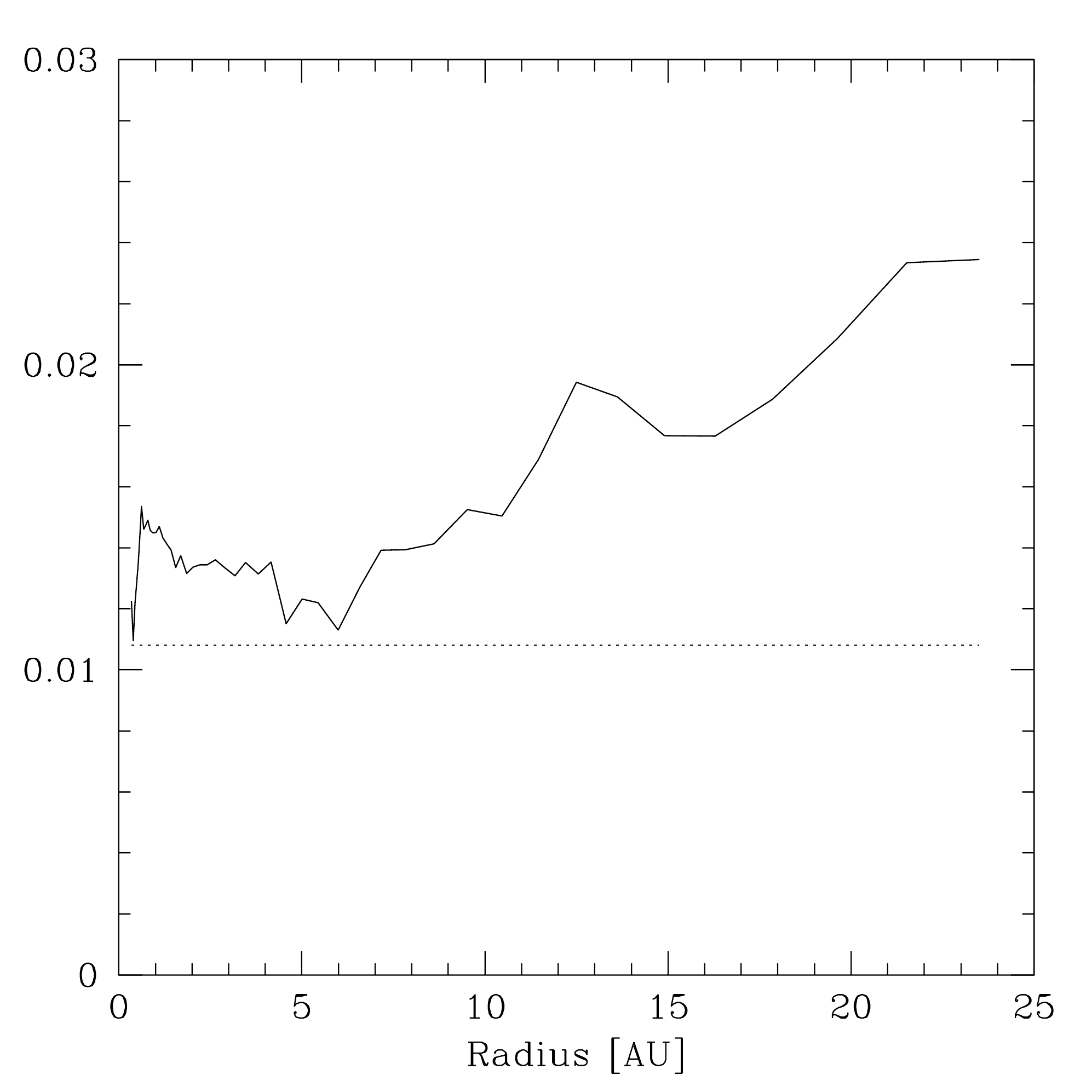}
\includegraphics[width=1.0\columnwidth]{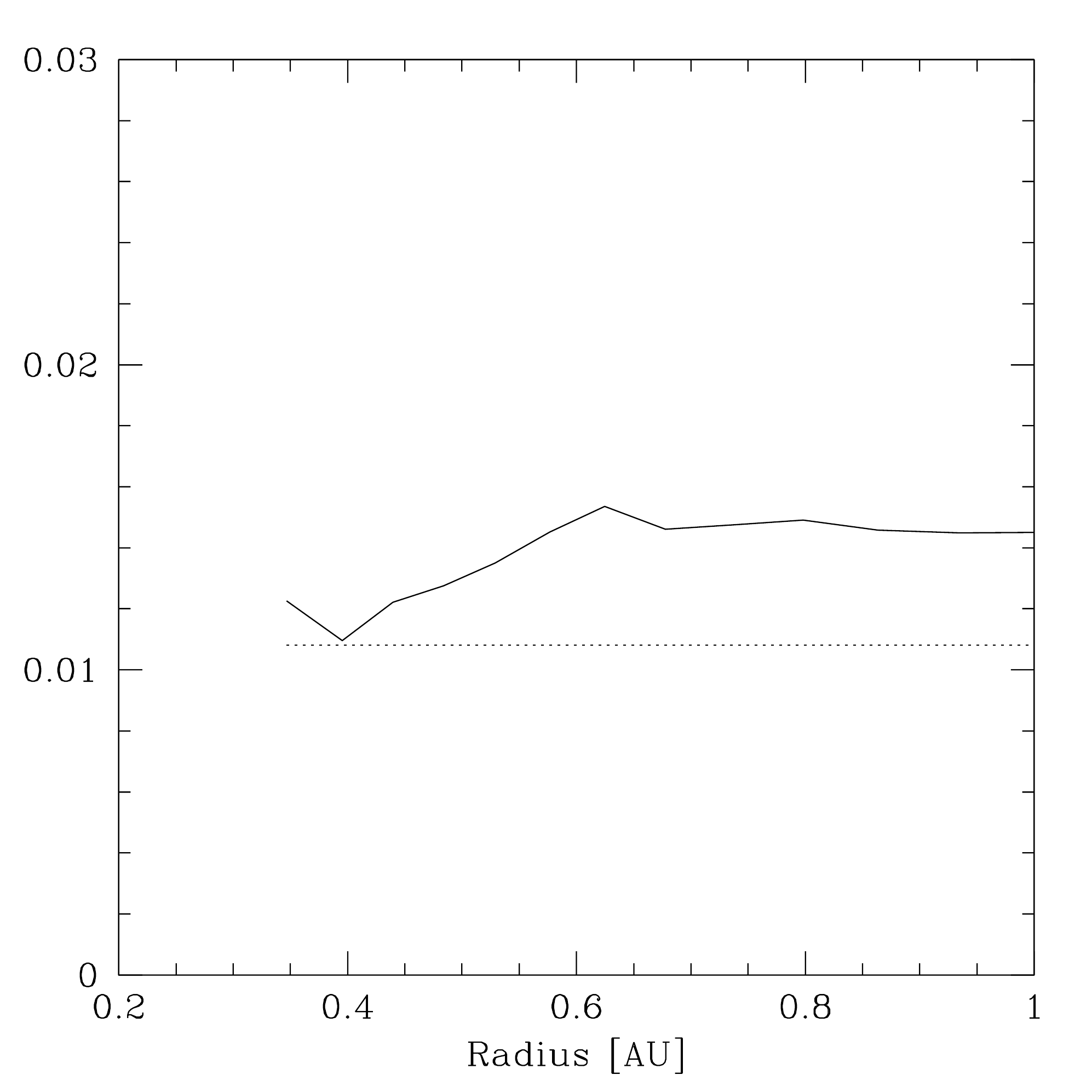}
  \caption{Plot of disc aspect ratio, H/R (solid line), against the RHS of equation~\ref{eq:analytics_ii} (dotted line) for simulation p2-beta3 for the radial range of the entire disc (upper panel) as well as zoomed into the inner regions (lower panel).  Condition~\ref{eq:analytics_ii} is marginally satisfied at $\approx 0.4$AU where the disc first fragments, confirming the analytical predictions in Section~\ref{sec:analytics}.}
  \label{fig:H_R_S-2}
\end{figure}

\begin{figure}
  \includegraphics[width=1.0\columnwidth]{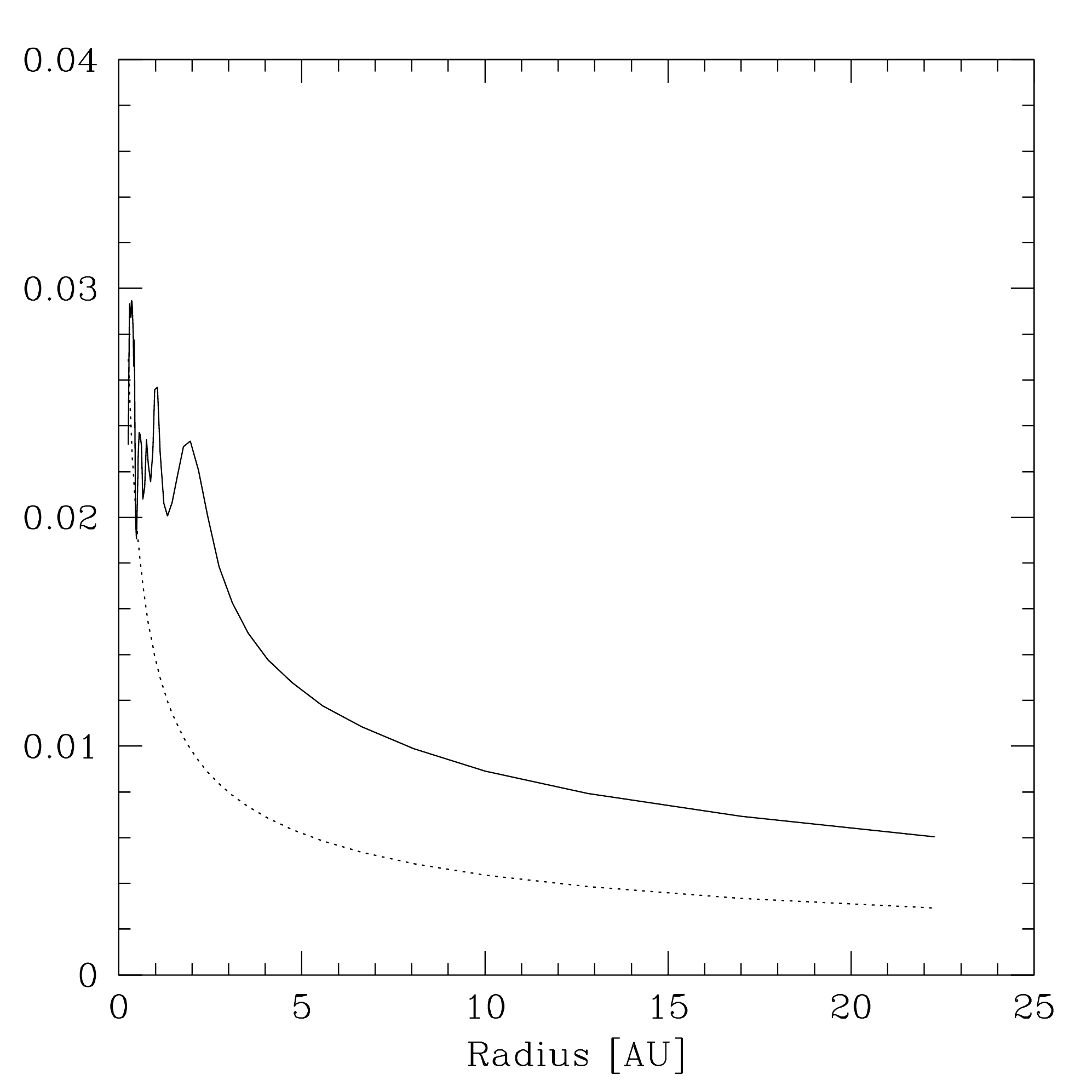}
  \includegraphics[width=1.0\columnwidth]{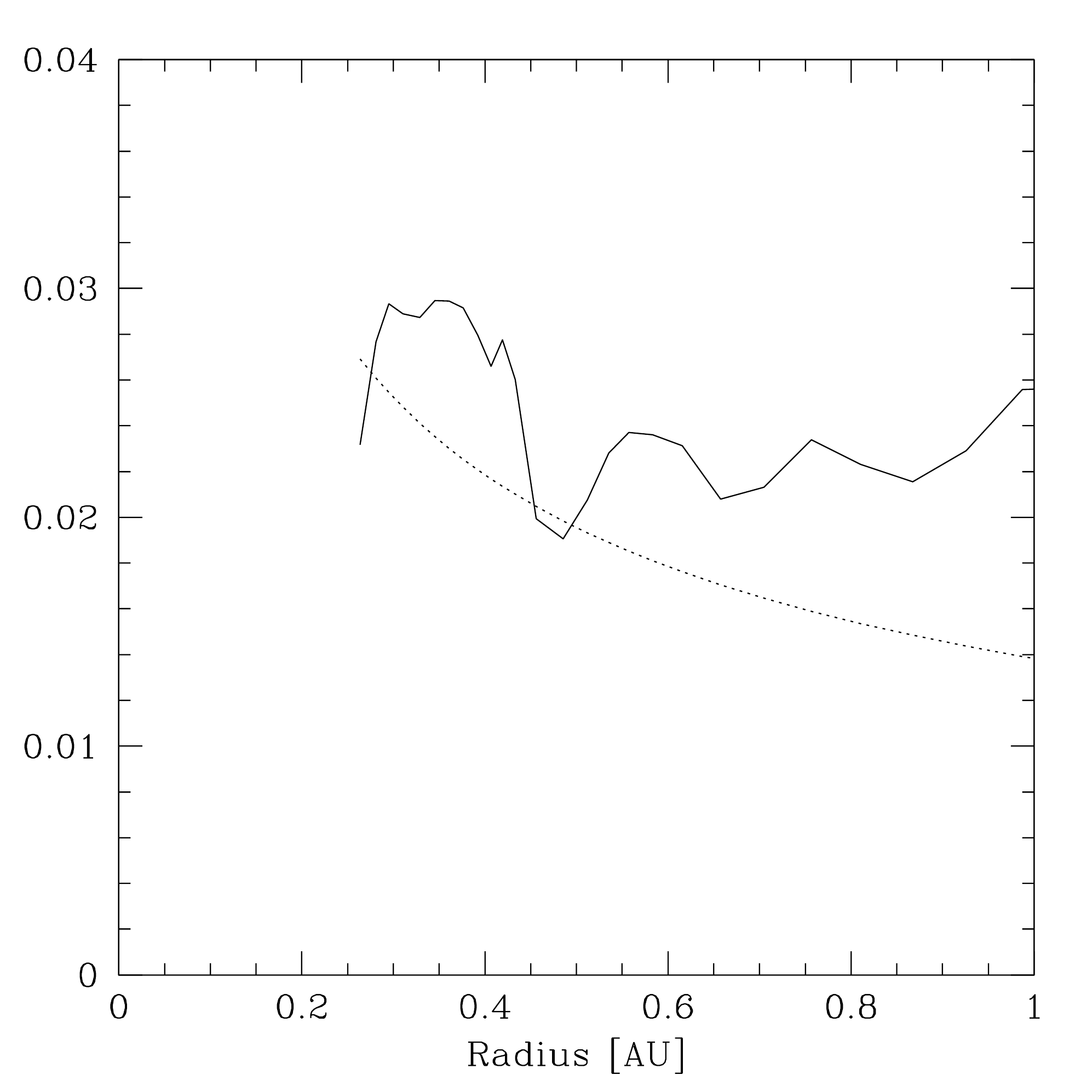}
  \caption{Plot of disc aspect ratio, H/R (solid line), against the RHS of equation~\ref{eq:analytics_ii} (dotted line) for simulation p2.5-beta3.5 for the radial range of the entire disc (upper panel) as well as zoomed into the inner regions (lower panel).  Condition~\ref{eq:analytics_ii} is satisfied at $\approx 0.4$AU where the disc first fragments, confirming the analytical predictions in Section~\ref{sec:analytics}.}
  \label{fig:H_R_S-2.5}
\end{figure}

The analytical theory for $p \gtrsim 2$ suggested that if the disc was to fragment, it would do so in the inner regions of the disc.  Figures~\ref{fig:S-2} and~\ref{fig:S-2.5} show that this is indeed the case.  Figure~\ref{fig:H_R_S-2} shows that the analytical condition is just satisfied for simulation p2-beta3 at the location at which the fragment forms.  Figure~\ref{fig:H_R_S-2.5} shows that for simulation p2.5-beta3.5, the analytical condition is also satisfied at the location at which fragmentation occurs soon after.

We therefore show that as the surface mass density profile is steepened so that more of the mass is concentrated in the inner regions of the disc, fragmentation moves towards smaller radii.  It is important to note that the trend that fragmentation moves to smaller radii for steeper surface mass density profiles is valid even when considering a uniform value of $\beta$ (compare simulations Reference-beta3, p1.5-beta3, p2-beta3 and p2.5-beta3 which are run using $\beta = 3$ and fragment at $\approx 8$, 1.7, 0.45 and 0.3~AU, respectively).

In addition, the results summarised in Table~\ref{tab:main_sim} show that a single value of $\beta_{\rm crit}$ is not applicable over all surface mass density profiles since the minimum value of $\beta$ that a disc can have without fragmenting varies with the surface mass density profile.

Simulation p2.5-beta3.5-Q5 was the same as simulation p2.5-beta3.5 but had an initial disc that was hotter by a factor of 25/4.  The results of this simulation show that the disc still fragmented in the inner regions.

\subsection{Effect of the cooling timescale, $\beta$, on the fragment location}
\label{sec:beta_Rf}

In Section~\ref{sec:analytics} the analytical work presented suggested that for a fragmenting disc with a shallow surface mass density profile ($p<2$), a decrease in the value of $\beta$ would cause the location at which the first fragment forms to move inwards to smaller radii.

\begin{figure}
  \includegraphics[width=1.0\columnwidth]{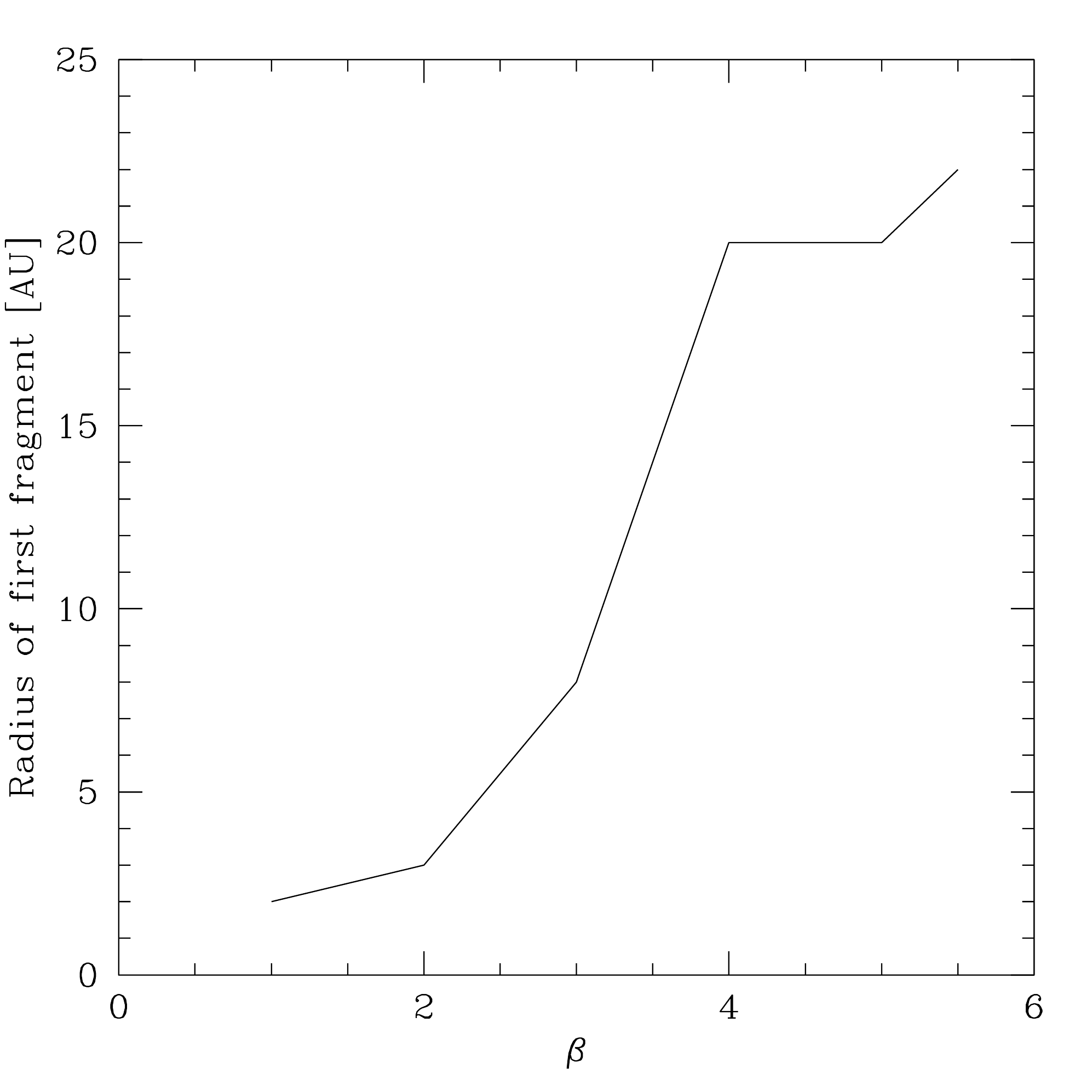}
  \caption{The radius at which the first fragment forms in the Reference simulations.  The discs in these simulations are identical with a surface mass density profile, $p=1$, but were run with different values of the cooling timescale in units of the orbital timescale, $\beta$.  The radius at which the first fragment forms moves inwards with more efficient cooling.}
  \label{fig:beta_Rfrag}
\end{figure}

Figure~\ref{fig:beta_Rfrag} shows the radius at which the first fragment forms for different values of $\beta$ (simulations Reference-beta5.5, Reference-beta5, Reference-beta4, Reference-beta3, Reference-beta2 and Reference-beta1).  We can see a clear trend showing that the radius of fragmentation moves inwards for more efficient cooling.

\subsection{The influence of star mass on fragmentation}
\label{sec:results_mstar}
In Section~\ref{sec:analytics}, we showed that decreasing the star mass is more likely to cause conditons~\ref{eq:analytics_i} and~\ref{eq:analytics_ii} to be satisfied over a larger part of the disc and hence the disc is more likely to fragment.  We test three identical discs with star masses of 0.5, 1 and 2 $\rm M_{\odot}$ (simulations p1-Mstar0.5, Reference-beta5 and p1-Mstar2, respectively) which are run with the same cooling timescale, $\beta = 5$.  It can be seen from Table~\ref{tab:main_sim} that the Reference-beta5 disc first fragments at $R_{\rm f} \approx 20$~AU.  However, when the star mass is halved, the first fragment forms at $R_{\rm f}\approx 13$~AU.  Since the RHS of condition~\ref{eq:analytics_ii} $\propto R^2/M_{\star}$, if the star mass is halved for the same value of $\beta$ (and hence the same value of the LHS of condition~\ref{eq:analytics_ii}), the radius at which the first fragment forms, $R_{\rm f}$, decreases by a factor of $\sqrt 2$.  Conversely, doubling the star mass makes it harder for the condition to be satisfied and indeed the disc in simulation p1-Mstar2 does not fragment.  Instead it settles into a state of marginal stability with $Q \approx 1$.

\subsection{The influence of disc mass on fragmentation}
\label{sec:results_mdisc}
The analytics presented in Section~\ref{sec:analytics} showed that increasing the disc mass (and hence increasing $\Sigma_o$) allows conditons~\ref{eq:analytics_i} and~\ref{eq:analytics_ii} to be satisfied over a larger part of the disc and hence the disc is more likely to fragment.  We initially test this by comparing the results of simulations p1-Mdisc0.05, Reference-beta5 and p1-Mdisc0.2 which are identical discs except that the disc masses are $0.05 \rm M_{\odot}$, $0.1 \rm M_{\odot}$ and $0.2 \rm M_{\odot}$, respectively.  Table~\ref{tab:main_sim} shows that doubling the disc mass from $0.1 \rm M_{\odot}$ to $0.2 \rm M_{\odot}$ does indeed cause the fragmentation conditions derived in Section~\ref{sec:analytics} to be satisfied over a larger portion of the disc since the first fragments form at $R_{\rm f} \approx 20$~AU and $\approx 14$~AU, respectively.  Since the RHS of condition~\ref{eq:analytics_ii} $\propto \Sigma R^2$, if the disc mass (and hence $\Sigma$) is doubled for the same value of $\beta$ (and hence the same value of the LHS of condition~\ref{eq:analytics_ii}), the radius at which the first fragment forms, $R_{\rm f}$, decreases by a factor of $\sqrt 2$.  However, halving the disc mass makes it harder for the conditions to be satisfied and consequently, the disc does not fragment.

In addition, we also simulate a very low mass disc ($M_{\rm disc} = 0.01 \rm M_{\odot}$) and found that it fragments if $\beta=1$, 2 and 2.5 but not for $\beta=3$ (simulations p1-beta1-Mdisc0.01, p1-beta2-Mdisc0.01, p1-beta2.5-Mdisc0.01 and p1-beta3-Mdisc0.01, respectively).  As $\beta$ increases, the fragment location moves out in the disc, as found in Section~\ref{sec:beta_Rf}.   It is clear that, as with varying the star mass, the disc mass plays a crucial role in the fragmentation and the condition for fragmentation cannot simply be described using a single critical value of the cooling timescale.

\begin{figure} \includegraphics[width=1.0\columnwidth]{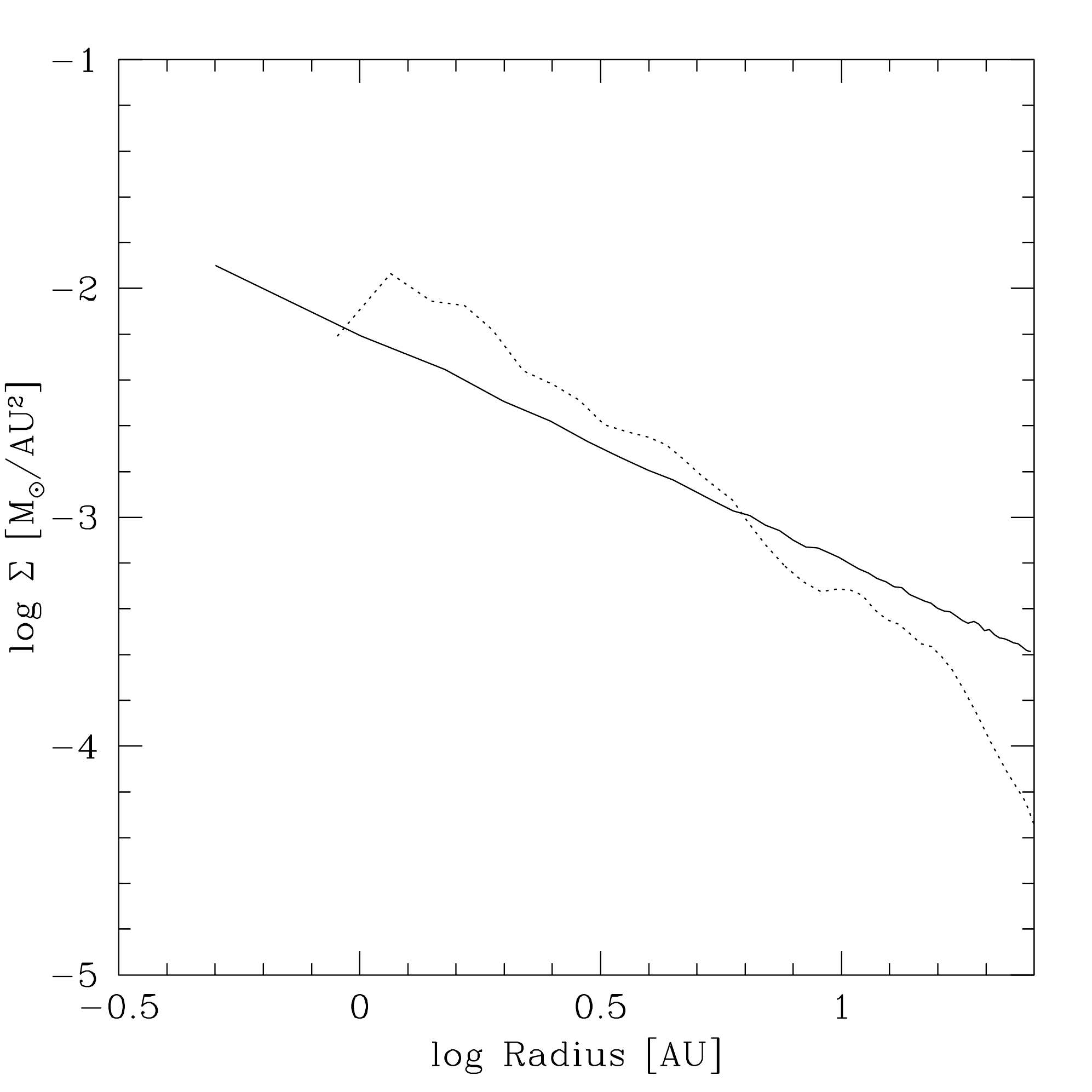}
  \caption{Surface mass density profiles for simulation p1-beta7-Mdisc1 at the start (solid line) and at a time more than 4~ORPs later (dotted line).  Unlike the low mass simulations whose surface mass density profiles do not change throughout the simulations, the profile for this disc steepens causing a change in the effective values of $\Sigma_o$ and $p$.}
  \label{fig:steepening_sigma}
\end{figure}

\begin{figure}
  \includegraphics[width=1.0\columnwidth]{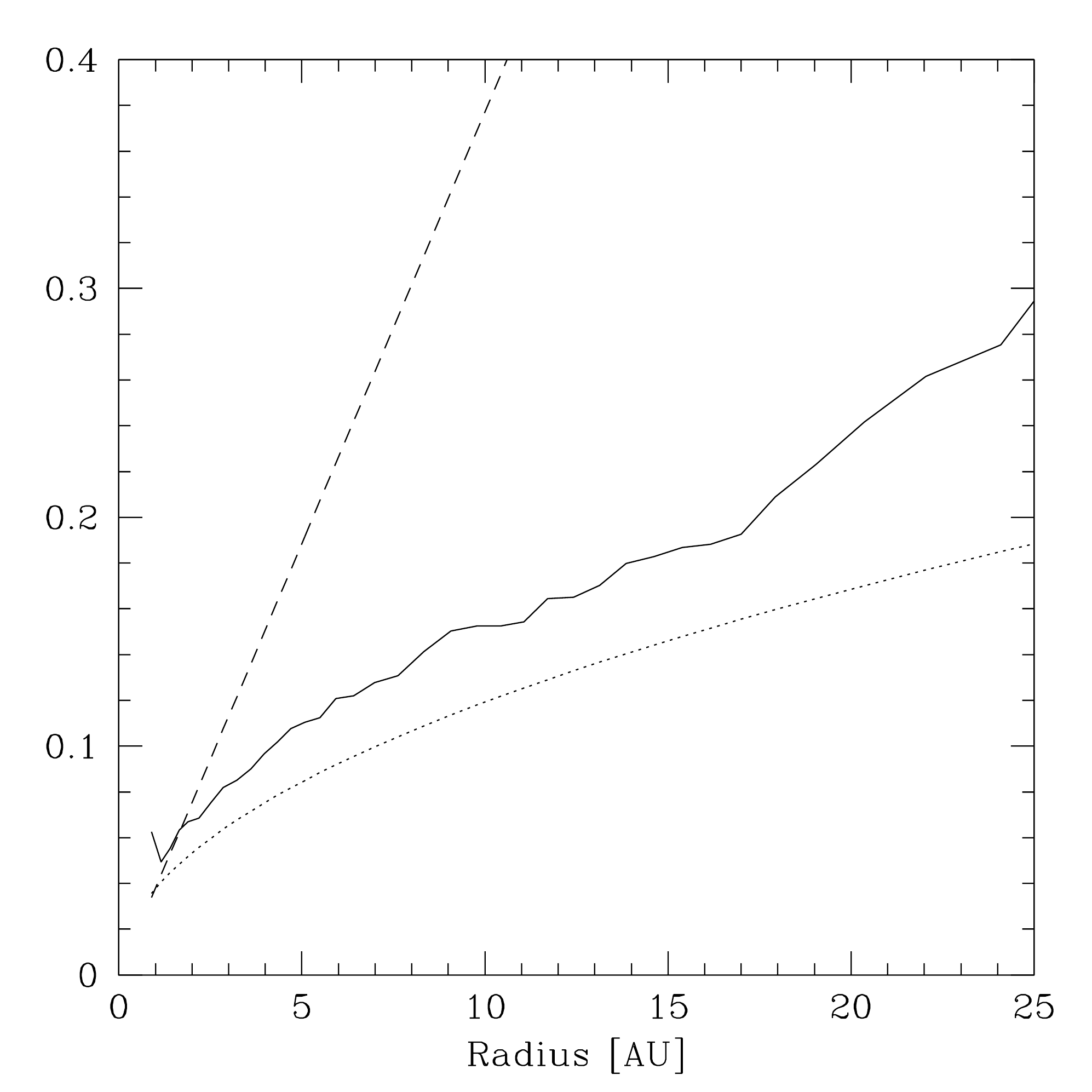}
  \caption{Plot of disc aspect ratio, H/R (solid line), for simulation p1-beta7-Mdisc1, against the RHS of equation~\ref{eq:analytics_ii} using the initial values of $\Sigma_o$ and $p$ (dashed line) and the new values of $\Sigma_o$ and $p$ determined after the disc has evolved for $> 4 \rm ORPs$ by which time its surface mass density profile has changed.  The condition is satisfied using the initial values of $\Sigma_o$ and $p$ but not using the new values and hence the disc does not fragment.}
  \label{fig:H_R_steepening_sigma}
\end{figure}

In addition, we simulate higher mass discs, $M_{\rm disc} = 0.3 \rm M_{\odot}$ and 0.5, which are run using $\beta = 8$ and 10 respectively (simulations p1-beta8-Mdisc0.3 and p1-beta10-Mdisc0.5) as well as discs with $M_{\rm disc} = 1 \rm M_{\odot}$ which we simulate using $\beta = 5$, 10 and 15 (simulations p1-beta5-Mdisc1, p1-beta10-Mdisc1 and p1-beta15-Mdisc1, respectively).  We find that with the exception of simulation p1-beta5-Mdisc1, the discs do not fragment.  Figure~\ref{fig:steepening_sigma} shows the surface mass density profile of simulation p1-beta7-Mdisc1 at the start and more than 4~ORPs after the start of the simulation.  It can be seen that unlike the lower mass discs, the profile steepens and the value of $\Sigma_o$ increases.  This is the case for all the non-fragmenting high mass disc simulations.  Figure~\ref{fig:H_R_steepening_sigma} shows the plot of the aspect ratio profile of this simulation against the RHS of condition~\ref{eq:analytics_ii} and shows that condition~\ref{eq:analytics_ii} is just satisfied in the inner regions.  However, since during the simulation the disc mass redistributes itself, the surface mass density profile changes and consequently, using the newly obtained values of $\Sigma_o$ and $p$, condition~\ref{eq:analytics_ii} is not satisfied.  We note that the only high mass disc that does fragment (simulation p1-beta5-Mdisc1), does so because the cooling time is so rapid that fragmentation occurs before the disc has had the chance to restructure itself.

\subsection{The role of the disc radius on fragmentation}
\label{sec:Rout}

\begin{figure} \includegraphics[width=0.95\columnwidth]{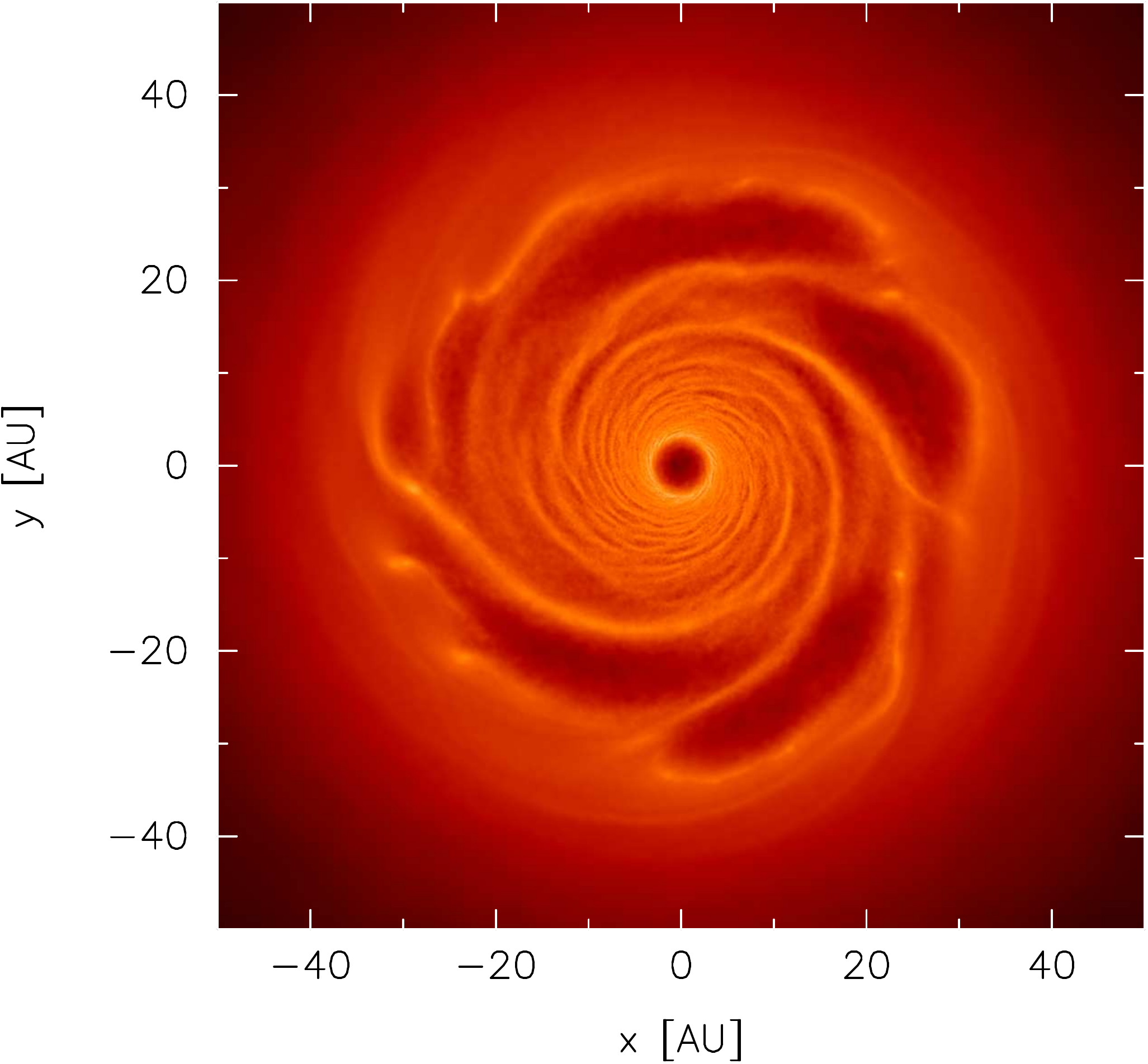}
  \caption{Surface mass density rendered image of the fragmenting disc in simulation p1-beta8-extended with initial surface mass density profile $\Sigma \propto R^{-1}$, but extending to 50~AU rather than 25~AU.  This simulation was run with $\beta=8$.  According to \citet{Rice_beta_condition}, this disc should not fragment since the cooling timescale $\beta$ is larger than the critical value previously obtained with a radius of 25~AU.  This simulation shows that the fragmentation criterion is more complex than a single critical cooling parameter.  The colour scale is a logarithmic scale ranging from log $\Sigma = -8$ (dark) to $-2$ (light) $\rm M_{\odot}/AU^2$.}
  \label{fig:b8_extended}
\end{figure}

In Section~\ref{sec:analytics}, we showed that for shallow surface mass density profiles ($p<2$), fragmentation might occur for \emph{any} value of the cooling timescale, if the disc is large enough.

Simulation Reference-beta6 (a disc with $R_{\rm out} = 25$~AU) did not fragment and though we did not run the same simulation with $\beta=7$ or 8, we would expect that they would also not fragment.  However, extended discs with the same values of $\Sigma_o$ and $p$ as Reference-beta6 do indeed fragment for $\beta=6$ (simulation p1-beta6-extended), $\beta=7$ (p1-beta7-extended) and $\beta=8$ (p1-beta8-extended; Figure~\ref{fig:b8_extended}) with the first fragments forming at $R_{\rm f}\approx 25$, 29 and 30~AU, respectively.  Similarly, we also simulate an extended disc with $p=1.5$ and $\beta=4$ (simulation p1.5-beta4-extended) and show that whilst the same disc truncated at $R_{\rm out} = 25$AU does not fragment, the extended disc does indeed fragment (at $R_{\rm f} \approx 33$AU).

In addition, given that in Section~\ref{sec:results_mdisc} we showed that the disc mass plays a part in whether fragmentation occurs or not, we simulate a $0.1\rm M_{\odot}$ disc which extends to $R_{\rm out} =50$AU (simulation p1-beta6-Mdisc0.1-extended).  The surface mass density profile is the same as in simulation p1-beta6-extended ($p=1$) but $\Sigma_o$ is decreased.  The results show that the disc fragments at $R_{\rm} \approx 40$AU (c.f. $\approx 25$~AU for simulation p1-beta6-extended).  Therefore, whilst the disc mass affects where in the disc the first fragment forms, the conclusion that a small disc which does not fragment for a particular value of $\beta$ may fragment at larger radii for the same value of $\beta$ is still valid.  Furthermore, we can see that if the disc to star mass ratio is kept constant at $M_{\rm disc}/M_\star = 0.1$ for an extended disc, the disc fragments (at $R_{\rm f} \approx 34$~AU; simulation p1-beta6-Mdisc0.2-Mstar2-extended) whilst the small disc with the same disc to star mass ratio does not fragment, further corroborates the fact that the radius of the disc is important.

\section{Discussion}
\label{sec:disc}

It has previously been accepted that for a self-gravitating disc whose only source of heating is internally generated from the gravitational instability, the disc will fragment if the cooling timescale is short enough \citep{Gammie_betacool}.  However, we find that fragmentation at a given radius is not only dependent on the cooling timescale, $\beta$, but also on the disc surface density (i.e. disc mass and profile) and the star mass.  

This is in contrast to \cite{Rice_beta_condition} who suggested that the fragmentation criterion is independent of the disc mass though in agreement with \cite{Rice_Gammie_confirm} who found that for a higher disc mass, fragmentation was easier: using $\beta = 5$, they found that a $0.25 \rm M_{\odot}$ disc fragmented whilst a $0.1 \rm M_{\odot}$ disc did not, but instead required a lower value of $\beta$.  In particular, we highlight that in the past, it has been thought that a massive disc is required for fragmentation to occur.  However, we show that it is indeed possible for low mass discs ($M_{\rm disc} \sim \rm O(0.01) \rm M_{\odot}$) to fragment if the cooling in the discs is rapid enough.  On the other hand, for high mass discs ($M_{\rm disc} \ge 0.3 \rm M_{\odot}$ within $R_{\rm out} = 25$~AU), the discs do not fragment, unless the cooling time is fast, due to a steepening of the surface mass density profile and an increase in $\Sigma_o$ making condition~\ref{eq:analytics_ii} harder to satisfy.  \cite{Lodato_Rice_massive_disc} also find that with larger disc masses, matter is redistributed causing the surface mass density profile to steepen.

In addition, we find that the critical value of $\beta$ found for one particular surface mass density profile is not applicable to another disc mass distribution.  Our simulations show that for a steeper surface mass density profile, the cooling timescale required for a disc to fragment is smaller.  \cite{Cossins_opacity_beta} found that for a disc with $\Sigma \propto R^{-3/2}$, $\beta_{\rm crit} \sim 4-4.5$ whereas \cite{Rice_beta_condition} found that $\beta_{\rm crit} \sim 6-7$ for a disc with $\Sigma \propto R^{-1}$.  In addition,  \cite{Rice_Gammie_confirm} found that for a surface mass density profile of $\Sigma \propto  R^{-7/4}$, the fragmentation boundary for a $0.1 \rm M_{\odot}$ disc around a $1 \rm M_{\odot}$ star is between $\beta = 3$ and 5 which \cite{Rice_beta_condition} note is inconsistent with their results.  However, our results explain these previous inconsistencies present in the literature.

We also find that for $p<2$, if a disc does not fragment for a particular cooling timescale in units of the orbital timescale, $\beta$, a larger disc with the same surface mass density profile may well fragment (compare simulations Reference-beta6 and p1-beta6-extended or p1.5-beta4 and p1.5-beta4-extended, and also p1-beta7-extended or p1-beta8-extended).  Therefore, a critical cooling timescale can only be specified for a particular surface mass density at a particular disc radius and can therefore not be a general rule.  The previous fragmentation criterion found by \cite{Rice_beta_condition} was for a disc to star mass ratio of 0.1 with $M_{\star} = 1 \rm M_{\odot}$, a surface mass density profile of $\Sigma \propto R^{-1}$ and $R_{\rm out} = 25$~AU.

Finally, we also show that the mass of the star plays a part in whether a disc fragments or not.  Therefore, as shown by the RHS of condition~\ref{eq:analytics_ii}, whether a disc fragments or not is essentially a ``trade-off'' between the local surface mass density and the star mass.

\subsection{The link between $\beta$, $M_{\rm disc}$, $M_{\star}$, the surface mass density profile, $p$, and fragmentation}

\begin{figure}
  \includegraphics[width=1.0\columnwidth]{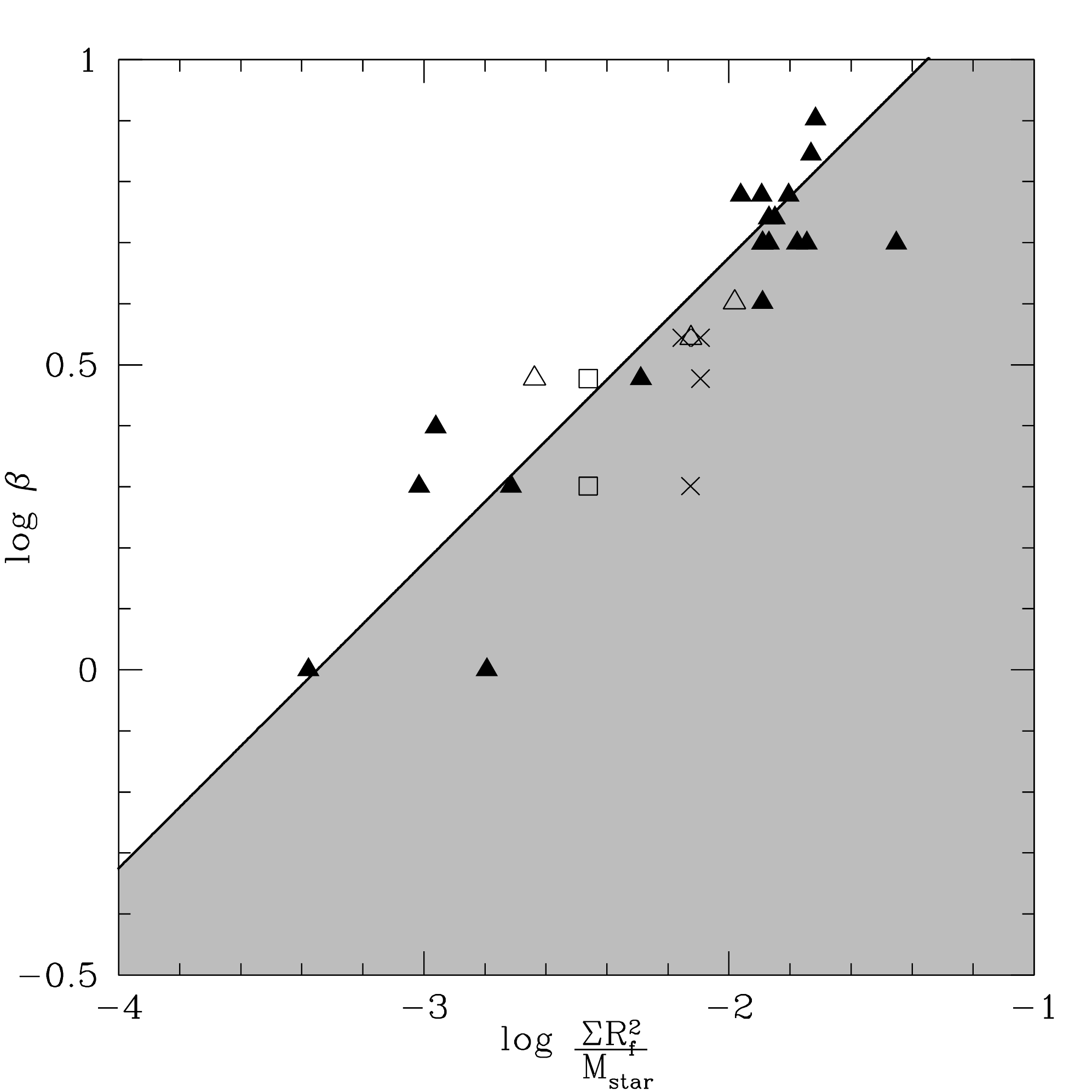}
  \caption{Logarithmic graph showing the trend between $\beta$ and $\Sigma R_{\rm f}^2/M_{\star}$ determined by considering the location at which the first fragment forms in the discs, $R_{\rm f}$.  The results include those simulations with a surface mass density profile, $p=1$ (filled triangles), $p=1.5$ (open triangles), $p=2$ (open squares) and $p=2.5$ (crosses).  It is clear that a single critical value of $\beta$ is not the case for all discs and that there is a relation between $\beta$, $M_{\rm disc}$, $M_{\star}$ and the surface mass density profile, $p$, that determines whether fragmentation occurs or not.  The trendline has been determined by considering discs with shallow surface mass density profiles, $p<2$ only as those discs with $p \gtrsim 2$ will always fragment in the innermost regions first.  The grey shaded region is where we expect subsequent fragmentation may take place in discs with $p<2$.}
  \label{fig:the_big_link}
\end{figure}

Section~\ref{sec:results} shows that the fragmentation criterion is clearly a complex problem which cannot simply depend on a single critical cooling timescale as has previously been thought to be the case.  Equation~\ref{eq:analytics_ii} and the results presented here clearly show that there is a link between the cooling timescale (in terms of $\beta$), the disc mass (or more accurately, the local surface mass density) and the star mass.  

Such a link can be explained physically.  As a disc cools, gravitational instability develops resulting in density fluctuations above and below the unperturbed density, $\delta \rho / \rho$.  The spiral structures involve shocks which produce heat that may balance the disc's cooling, thus reaching an equilibrium state.  If the disc mass was irrelevant, for any particular star mass, and $\beta$, one would expect the fluctuations, $\delta \rho / \rho$, to be the same in all discs with the same surface density profile.  However, comparing a low mass disc with a high mass disc with the same relative density fluctuations ($\delta \rho / \rho$), the density enhancement, $\delta \rho$, in the higher-mass disc will clearly be greater.  At some disc mass, this enhancement will be self-gravitating (i.e. it will be a fragment), while in a lower-mass disc the density fluctuation will not form a fragment (unless the value of $\beta$ is lowered).  Similarly, if the disc is kept the same and the star's mass is increased, a given density enhancement may be sheared apart by the differential rotation so that a low value of $\beta$ will be required for fragmentation.

Figure~\ref{fig:the_big_link} shows a graph of $\beta$ against $\Sigma R_{\rm f}^2/M_\star$, where $R_{\rm f}$ is the radius at which the first fragment forms, and includes all the fragmenting simulations presented in this paper (including the Benchmarking simulations).  When interpreting these results, one should also note that an increase in $\Sigma R^2 / M_\star$ does not necessarily imply an increased disc to star mass ratio: it is possible to have a low disc to star mass ratio but using a steeper surface mass density profile.  We can see that there is a clear trend that as the RHS of equation~\ref{eq:analytics_ii} increases, so too does the value of $\beta$ that will allow a fragment to form.  The trendline presents the relation:

\begin{equation}
\beta = \eta \bigg(\frac{\Sigma R_{\rm f}^2}{M_{\star}}\bigg)^{\delta},
\end{equation}
where $\delta \approx 1/2$ and the constant of proportionality, $\eta \approx 47$, which we find using a least squares fit.  We have performed a few calculations with higher resolution ($2\times 10^6$ SPH particles) and find that this trend of $\beta$ increasing with $\Sigma R^2 / M_\star$ is maintained, although the exact values of $\eta$ and $\delta$ may change slightly if we were able to perform all the calculations with high resolution.

It is also important to note that the trendline has only been produced using the results of simulations with $p<2$.  This is because for $p \gtrsim 2$, the disc will always fragment in the innermost regions first.  Consequently, if the results for $p \gtrsim 2$ were included, this would cause the trendline to be somewhat skewed. 

The trendline can give some very useful information for those simulations with $p<2$.  The grey region is where we predict subsequent fragmentation is feasible.  Traversing the plot in a vertical direction downwards from the trendline into the grey region, at any particular value of $\Sigma R^2 /M_{\star}$ the disc will fragment at all values of $\beta$ less than the limit given by the trendline (though the radius being considered will not necessarily be the first location at which fragmentation occurs).  Similarly, traversing the plot in a horizontal direction from the trendline to the right side of the plot into the grey region, for a particular value of $\beta$, fragmentation is possible at a particular radius if the disc mass is increased or star mass is decreased.  Similarly,  \emph{for a particular value of $\beta$ and any one combination of the disc to star mass ratio}, fragmentation is also possible at larger radii than the location specified by the trendline i.e. to the right hand side of the trendline.  Therefore, the trendline predicts the minimum possible radius at which fragments could in theory form in discs with shallow surface mass density profiles, $p<2$.

For discs with $p \gtrsim 2$ which fragment in the inner regions, we expect that subsequent fragmentation may take place further out in the disc as far out as given by the trendline in Figure~\ref{fig:the_big_link}.  In other words, for these discs, we expect there to be a maximum radius outside of which fragmentation will not occur (since the surface mass density fall-off is steep, the outer regions of these discs may struggle to have enough mass for gravitational instability to be significant).  However, since we stop the simulations soon after the first fragment forms due to the increased computational resources required to follow the simulations further, we are unable to test this prediction.  Further work needs to be done in this area and is beyond the scope of this paper.  We also note that in real discs, for fragmentation to occur in the inner regions, the cooling time would have to be very small since the dynamical time at small radii would also be very small.  Such short cooling times may not be possible in real discs.

It is also important to note that these simulations have been carried out using a ratio of specific heats, $\gamma = 5/3$.  As shown by \cite{Rice_beta_condition}, the ratio of specific heats plays a key role in the fragmentation boundary.  We therefore also anticipate a further dependency on the equation of state.  Furthermore, for high mass discs ($M_{\rm disc} \gtrsim 0.3 \rm M_{\odot}$), we find that the initial surface mass density conditions (i.e. $\Sigma_o$ and $p$) cannot be used to determine whether the disc will fragment or not.  This is because as the disc evolves, the surface mass density profile steepens causing $\Sigma_o$ and $p$ to change.  Consequently, though parts of the disc may start off in the grey shaded region of Figure~\ref{fig:the_big_link} and hence may be expected to fragment, the disc may restructure itself on a timescale faster than the cooling timescale such that it moves out of this region and hence does not fragment.

Following the work of \cite{Gammie_betacool} and \cite{Rice_beta_condition}, a number of authors have used the critical cooling timescale, $\beta_{\rm crit}$ (or gravitational stress, $\alpha_{\rm GI, max}$) to predict fragmentation in realistic discs \citep[e.g.][]{Clarke2009_analytical,Rafikov_SI,Cossins_opacity_beta,Kratter_runts}.  In light of the new results presented here, we would encourage previous conclusions based on these critical values to be revisited.

\section{Conclusions}
\label{sec:conc}

We present an analytical approach to examine the fragmentation of self-gravitating protoplanetary discs, and confirm the results using global three-dimensional numerical simulations.  Our key result is that fragmentation does not simply depend on the disc cooling timescale, $\beta$, but also on the ratio of the surface mass density at radius, $R$, to the stellar mass, i.e. $\Sigma R^2 / M_{\star}$.  We find that fragmentation occurs when

\begin{equation}
\beta < \eta \bigg(\frac{\Sigma R_{\rm f}^2}{M_{\star}}\bigg)^{\delta},
\label{eq:beta_lt}
\end{equation}
where $\delta \approx 1/2$ and $\eta \approx 47$.  For a power-law surface mass density, $\Sigma \propto R^{-p}$, this relation predicts the innermost radii at which subsequent fragments can form in a disc with shallow surface mass density profiles ($p<2$) as well as the radius \emph{outside of which fragments cannot form} in discs with steep surface mass density profiles, ($p \gtrsim 2$).  Generally, we find that an increase in the steepness of the disc surface mass density profile promotes fragmentation at smaller radii.

\section*{Acknowledgments}
We thank the anonymous referee for their thoughtful comments on the paper.  The calculations reported here were performed using the University of Exeter's SGI Altix ICE 8200 supercomputer and Intel Nehalem (i7) cluster.  The disc images were produced using SPLASH \citep{SPLASH}.  MRB is grateful for the support of a EURYI Award which also funded FM. This work, conducted as part of the award `The formation of stars and planets: Radiation hydrodynamical and magnetohydrodynamical simulations' made under the European Heads of Research Councils and European Science Foundation EURYI (European Young Investigator) Awards scheme, was supported by funds from the participating organizations of EURYI and the EC Sixth Framework Programme.

\bibliographystyle{mn2e}
\bibliography{allpapers}

\end{document}